\documentclass[aps,prb,reprint,amsmath,amssymb,superscriptaddress,floatfix,letterpaper]{revtex4-2}
\usepackage{graphicx}
\usepackage{bm}
\usepackage{siunitx}
\makeatletter
\@ifclassloaded{revtex4-2}{}{%
  \def\p@subsection{\thesection\,}%
  \RenewDocumentEnvironment{ruledtabular}{}{\par\centering\let\rt@tabular\tabular\let\rt@endtabular\endtabular
    \def\tabular##1{\csname tabular*\endcsname{\linewidth}{@{\extracolsep{\fill}}##1}\hline\hline}%
    \def\endtabular{\hline\hline\csname endtabular*\endcsname}}{\par}}
\makeatother
\usepackage{floatsetup}
\usepackage[hidelinks]{hyperref}

\begin{document}

\title{Tunneling governs the current jumps of the boundary-driven XXZ chain}

\author{Rohit Narula}
\email{rnarula@physics.iitd.ac.in}
\affiliation{Department of Physics, Indian Institute of Technology Delhi, New Delhi 110016, India}

\date{\today}

\begin{abstract}
The boundary-driven spin-1/2 XXZ chain has an exactly known steady state. Its spin current jumps at the roots of unity, where the anisotropy $\Delta$ is the cosine of $\pi$ times a ratio of integers. At bath coupling $\varepsilon=2$, the current of a long chain falls by \qty{5.69}{\percent} as soon as the anisotropy leaves $\Delta=1/2$. The jump comes from the transfer matrix of the steady state, whose couplings repeat with a phase that drifts slowly along the matrix. At the root of unity this phase is frozen where the largest eigenvalue is smallest. Nearby it runs through all values, a larger eigenvalue takes over and the current drops. A finite chain keeps the current of the root of unity up to a crossover length, which we obtain as a semiclassical tunneling time. With no fitted parameter, this length matches the computed values at $\Delta=1/2$ to within a relative $\num{1.7e-4}$. For odd denominators and strong enough coupling, the crossover sharpens into a step. At strong driving only odd denominators keep a jump. For small $\eta=\arccos\Delta$ its size decays as $e^{-8G/\eta}$ up to a power-law prefactor, where $G$ is Catalan's constant and $8G$ a tunneling action.
\end{abstract}

\maketitle

\section{Introduction}

Transport coefficients of a many-body system are ordinarily smooth functions of its parameters, and a discontinuity in them marks a phase transition. The spin-1/2 XXZ chain is an exception, its spin Drude weight at nonzero temperature depending discontinuously on the anisotropy within a single phase. The Drude weight is defined as the weight of the zero-frequency component of the conductivity. Kohn related the weight of free acceleration in the low-frequency conductivity of a ring at zero temperature to the response of its energy to a flux threaded through it, a weight that vanishes in an insulator~\cite{kohn1964}, and on a lattice the weight is a stiffness, given by the response of the energy to a twist of the boundary condition~\cite{shastry1990}. A Drude weight remaining finite at nonzero temperature was conjectured to be generic to integrable systems~\cite{castella1995}, and conservation laws bound it from below~\cite{zotos1997} through the inequality of Mazur~\cite{mazur1969}, which bounds the Drude weight by the overlap of the current with conserved quantities. At high temperature this bound is a fractal function of the anisotropy~\cite{prosen2011open}, and it is improved by a continuous family of quasilocal conservation laws, conserved operators whose densities decay exponentially rather than vanish beyond a finite range~\cite{prosen2013quasilocal}. The Drude weight itself is obtained from the Bethe ansatz~\cite{zotos1999} and from exact thermodynamic Bethe-ansatz formulas that agree with generalized hydrodynamics, the hydrodynamics of systems with infinitely many conservation laws~\cite{urichuk2019}, with both it and the asymptotic rates of the spin current depending on the anisotropy in a completely discontinuous way~\cite{ilievski2017}. These quasilocal conserved charges are reviewed in Ref.~\cite{ilievski2016quasilocal}, and transport in integrable chains is covered in the lecture notes of Ref.~\cite{sirker2020} and in the review of Ref.~\cite{bertini2021rmp}.

In the closed chain the same sensitivity, to whether the anisotropy is the cosine of a rational multiple of $\pi$, shows up in time and frequency. For almost all easy-plane anisotropies the low-frequency conductivity diverges as the inverse square root of the frequency~\cite{agrawal2020}. An earlier account places a diffusive channel in parallel with the ballistic one~\cite{sirker2009}. The dynamics of the magnetization at the isotropic point is described by the Kardar-Parisi-Zhang scaling function~\cite{ljubotina2019}, with an effective diffusion constant growing as the cube root of time~\cite{gopalakrishnan2019}. Hydrodynamics with infinitely many conservation laws~\cite{castroalvaredo2016} yields the profiles of charges and currents that develop when two halves of the chain prepared in different states are joined~\cite{bertini2016}, and for the gapless chain whose two halves start at infinite temperature with opposite magnetizations the locally quasistationary state carrying the spin current is known analytically~\cite{collura2018}. Ultracold atoms in optical lattices realize Heisenberg chains with tunable anisotropy~\cite{jepsen2020}, making the anisotropy a control parameter of experiments on closed chains.

Beyond the closed chain, a current can also be imposed from the boundaries. With baths injecting spin at one end of the chain and removing it from the opposite end, transport becomes a property of a steady state. The chain then evolves under a quantum dynamical semigroup, the generators of which have a known general form~\cite{lindblad1976}, and it belongs to the large family of quantum systems driven dissipatively from their edges~\cite{landi2022rmp}. Matrix-product simulations of boundary-driven chains~\cite{prosen2009mps} found ballistic spin transport for easy-plane anisotropy in the XXZ chain~\cite{znidaric2011}. For maximal driving, by baths that only inject on one end and only remove on the opposite end, the steady state is an explicit matrix product for any bath coupling~\cite{prosen2011exact}. The matrix-product construction extends to a quadratic algebra, a set of relations obeyed by the matrices of the product, for the boundary-driven chain~\cite{karevski2013}, connected correlations and current fluctuations follow from it~\cite{buca2016}, and Ref.~\cite{prosen2015review} reviews these solutions.

The dependence of the current of the exact matrix-product solution on the anisotropy is, at present, known only in part. At a root of unity, where the anisotropy angle $\eta$ equals $\pi l/m$ with coprime integers $l$ and $m$, the transfer matrix that yields the expectation values of the solution reduces to $m\times m$, the current of the long chain being fixed by its leading eigenvalue~\cite{prosen2015review}. This formula yields the current separately at each root of unity, and the truncation of the transfer matrix on which it rests (Sec.~\ref{sec:model}) was noted with the solution~\cite{prosen2011exact}. What the formula does not give is the limit of the current along neighboring anisotropies, where the matrix has a different and growing size, and with it the size of the difference. Related discontinuities are on record. For a coherent boundary drive, the steady-state magnetization exhibits a fractal dependence on the interaction strength~\cite{arxiv2407}, and for the steady state considered here, at commensurate anisotropies, the finite-size corrections to the generating function of the local magnetization have been shown to decay with the block size at a rate that is a strongly discontinuous function of the anisotropy~\cite{manna2024}. The limit of strong dissipation, which we take in Sec.~\ref{sec:strong}, has been studied for the pure states it can prepare~\cite{popkov2017targeting} and for the spectrum of its Liouvillian~\cite{popkov2021zeno}. What is missing is quantitative, namely how large the discontinuities of the long-chain current are, at what length a finite chain shows them, and what law they follow as the denominator of the root of unity grows.

\begin{figure*}[t]
\includegraphics[width=\textwidth]{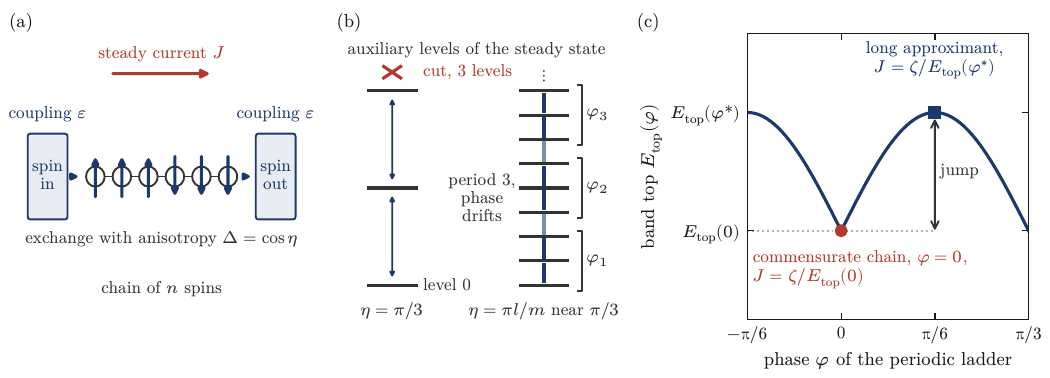}
\caption{\label{fig:schematic}Why the current of the driven chain jumps at a root of unity. (a)~A chain of $n$ spins with exchange anisotropy $\Delta=\cos\eta$. Baths of coupling $\varepsilon$ inject spin at the left end and remove it at the right (arrows sketch the spin pumped in and out), and a steady current $J$ flows. (b)~The auxiliary levels of the matrix-product steady state, the basis states of the space on which the matrices act. The transfer matrix couples each level to its neighbors, as the rungs of a ladder. For $\eta=\pi/3$ the ladder breaks off after three levels (red cross). At a nearby root of unity $\eta=\pi l/m$, an approximant, it has $m$ levels whose couplings repeat with period three while a phase $\varphi$ drifts from one block to the next (brackets $\varphi_1$, $\varphi_2$, $\varphi_3$, with the lighter coupling closing each block). (c)~The largest eigenvalue of the periodic ladder at a fixed phase, the band top $E_\mathrm{top}(\varphi)$ (blue curve, sketch), which is even in $\varphi$ and has the period $\pi/3$. The blocks of an approximant sample all phases, so its largest eigenvalue is set by the highest band top. The commensurate chain is at $\varphi=0$ (red circle), where the band top rises with a nonzero slope, a kink in the even function, and a long approximant attains the maximum at $\varphi^*=\pi/6$ (blue square), and the current, $\zeta$ divided by the band top, is different for the two. The higher band top of the approximant lowers its current by the jump (double arrow). The dotted line is the band top of the commensurate chain. The prefactor $\zeta$ is defined with Eq.~\eqref{eq:Jn}, and the dots above the right ladder indicate that the ladder continues to $m$ levels.}
\end{figure*}

In this paper, we determine for the maximally driven chain the size of the jumps in the long-chain current, the chain length at which they are resolved, and the law they obey at strong driving. The discontinuity itself is implicit in the exact solution. Its auxiliary space truncates at roots of unity~\cite{ilievski2014quantum}, and what we add is the size of the jump, the length that resolves it and its law. A jump here means that the current of a long chain at an anisotropy such as $\Delta=1/2$ differs by a finite amount from the current at anisotropies arbitrarily close to it, although neither the Hamiltonian nor the baths change abruptly and the steady state is unique. We find that the current of the infinite chain jumps at every root of unity. At the anisotropy $\Delta=1/2$ and bath coupling 2, the common limit of its neighbors lies \qty{5.69}{\percent} below that current. The reason is a phase on which the leading eigenvalue of the transfer matrix depends. In the chain at the root of unity, which we call the commensurate chain, this phase stays at its least favorable value, whereas at every nearby anisotropy it passes through all values (Fig.~\ref{fig:schematic}). A finite chain keeps the commensurate value until its length times its detuning from the root of unity passes a threshold that we give in closed form, and it crosses over at a length inversely proportional to that detuning. We call a nearby root of unity with a larger denominator an approximant. For the approximants closest to a root of unity this length is between $\num{5.6}$ and 97 times their denominator in the cases computed (Table~\ref{tab:nstar}). This length is derived as a tunneling time in the auxiliary space, the space on which the matrices of the matrix product act (Sec.~\ref{sec:tunnel}), and we find it in agreement with the computed lengths without a fitted parameter. Under strong driving only the jumps at odd denominators remain, and they decay exponentially with the inverse of the angle $\eta$, with Catalan's constant in the exponent. Both facts are traced to the strong-driving operator written in momentum space, where the momenta coupled by the anisotropy form a ring that is cut whenever the denominator is even (Sec.~\ref{sec:parity}).

The exponential law arises as a tunneling effect of the kind known from Bloch electrons in a magnetic field. For Harper's equation~\cite{harper1955} the nature of the spectrum depends on whether the flux is rational or irrational~\cite{hofstadter1976}, at rational flux the total width of the bands shrinks as the inverse of the denominator~\cite{thouless1983}, and near the band edges the splittings are governed by semiclassical tunneling~\cite{wilkinson1984}. The transfer matrix of the driven chain at strong driving is a difference operator of a related kind, a product of two trigonometric factors where Harper's is a sum, with the angle $\eta$ playing the role of Planck's constant. In what follows, Section~\ref{sec:model} recalls the exact solution, Sec.~\ref{sec:finite} deals with the jump at finite driving, Sec.~\ref{sec:length} with the length that resolves it, Sec.~\ref{sec:strong} with the strong-driving limit, and Sec.~\ref{sec:discussion} with what a test would require. The derivations are collected in the Appendixes.

\section{Model and exact steady state}
\label{sec:model}

We consider a spin-1/2 chain of length $n$ with Hamiltonian $H$,
\begin{equation}
H=\sum_{j=1}^{n-1}\left[2\sigma^+_j\sigma^-_{j+1}+2\sigma^-_j\sigma^+_{j+1}+\Delta\,\sigma^{\mathrm{z}}_j\sigma^{\mathrm{z}}_{j+1}\right],
\label{eq:H}
\end{equation}
where $\sigma^\pm_j$ and $\sigma^{\mathrm{z}}_j$ denote the raising, lowering, and third Pauli operators at site $j$, and $\Delta=\cos\eta$ sets the anisotropy, taken in the easy-plane range $|\Delta|<1$ so that the angle $\eta$ is real. The density matrix $\rho$ evolves under the Lindblad equation\begin{equation}
\frac{d\rho}{dt}=-i[H,\rho]+\sum_{\nu=1}^{2}\left(2L_\nu\rho L_\nu^\dagger-\{L_\nu^\dagger L_\nu,\rho\}\right),
\label{eq:lindblad}
\end{equation}
for time $t$, where two jump operators $L_1=\sqrt{\varepsilon}\,\sigma^+_1$ and $L_2=\sqrt{\varepsilon}\,\sigma^-_n$ describe baths with coupling $\varepsilon$ (one injecting spin on the first site, the other extracting spin on the last site). This corresponds to the maximally driven chain solved in Ref.~\cite{prosen2011exact}, and we keep its normalization throughout.

The steady state is a matrix product whose norm and current are determined by powers of one tridiagonal transfer matrix. The matrices of the product act on a space with the auxiliary levels $|k\rangle$, $k=0,1,2,\dots$ as basis states. With a complex parameter $s$ fixed by $\tan(\eta s)=\varepsilon/(2i\sin\eta)$ for $\varepsilon\neq2\sin\eta$, where it has no finite solution and the current is taken as the limit, and of which any solution serves because the entries below do not change when $\eta s$ is shifted by $\pi$, the norm of the state on $n$ sites is the element $Z_n=\langle0|T^n|0\rangle$ of a tridiagonal transfer matrix $T$. The diagonal entries $T_{k,k}$, the entries $T_{k,k+1}$ above the diagonal, and the entries $T_{k+1,k}$ below the diagonal read\begin{equation}
\begin{aligned}
T_{k,k}&=2\left|\cos[\eta(s-k)]\right|^2,\\
T_{k,k+1}&=\left|\sin[(2s-k)\eta]\right|^2,\\
T_{k+1,k}&=\sin^2[(k+1)\eta].
\end{aligned}
\label{eq:T}
\end{equation}
The current $J$ through any bond, the expectation value of $i(\sigma^+_j\sigma^-_{j+1}-\sigma^-_j\sigma^+_{j+1})$, is then
\begin{equation}
J(n)=\zeta\,\frac{Z_{n-1}}{Z_n},
\label{eq:Jn}
\end{equation}
where the prefactor $\zeta$ is $\varepsilon\left|\cos(\eta s)\right|^2$ and the norm $Z_{n-1}$ corresponds to the chain one site shorter. The matrix $T$ describes a hopping process between neighboring auxiliary levels~\cite{prosen2011exact}, and $Z_n$ is the summed weight of the walks of $n$ steps that leave level 0 and return to it, these weights not being probabilities owing to the fact that $T$ does not conserve their sum. We verified Eqs.~\eqref{eq:T} and~\eqref{eq:Jn} against the Lindblad generator null vector for chains of 3 to 7 sites, and the largest relative difference in currents is $\num{3.3e-15}$ (Appendix~\ref{si:validation}). These chains reach the fourth auxiliary level. Entries at higher levels come from the exact solution.

At a root of unity, $\eta=\pi l/m$ with coprime integers $l$ and $m$, the element $T_{m,m-1}=\sin^2(m\eta)$ vanishes, and the transfer matrix reduces to its first $m$ levels, the invariant subspace of the reducible representation of the quantum group $U_q(\mathfrak{sl}_2)$ at a root of unity~\cite{ilievski2014quantum,prosen2014quasilocal,pereira2014exactly}. The levels $k\geq m$ cannot be reached from level 0, and $T$ acts as an $m\times m$ matrix. This matrix is similar to the symmetric tridiagonal (Jacobi) matrix with diagonal $d_k=T_{k,k}$ and off-diagonal $e_k=(T_{k,k+1}T_{k+1,k})^{1/2}$, whose entries are non-negative, and its largest eigenvalue $\tau_1$ therefore belongs to a positive eigenvector. Since that eigenvector overlaps level 0, $Z_n$ grows as $\tau_1^n$ and the current of the infinite chain is
\begin{equation}
J_\infty=\frac{\zeta}{\tau_1},
\label{eq:Jinf}
\end{equation}
independent of the length. Transport is thus ballistic in the easy-plane regime~\cite{prosen2011exact}, and Eq.~\eqref{eq:Jinf} is the form given in Ref.~\cite{prosen2015review}. Everything below follows from how $\tau_1$ depends on $l/m$. In what follows, we denote by $J_\infty(p/q)$ the current at the root of unity $\eta=\pi p/q$ with coprime integers $p$ and $q$, and $J_\infty(1/3)$ is therefore the current at $\Delta=1/2$.

\section{The jump at finite driving}
\label{sec:finite}

\subsection{Its size at anisotropy 1/2}

\begin{figure*}[t]
\includegraphics[width=\textwidth]{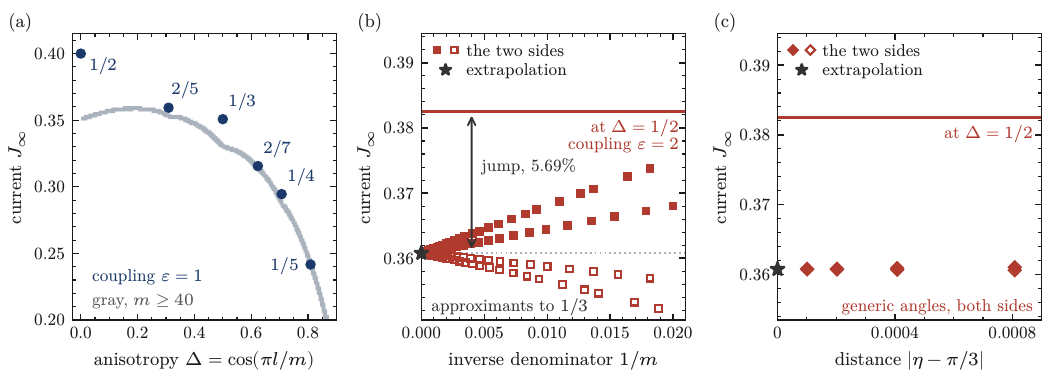}
\caption{\label{fig:jump}In the infinite chain, the current jumps at $\Delta=1/2$. (a)~$J_\infty$ at the roots of unity $\eta=\pi l/m$ for bath coupling $\varepsilon=1$, with gray dots for all denominators $40\leq m\leq200$ and blue circles for the low-denominator points labeled by $l/m$. (b)~At $\varepsilon=2$, the current along four sequences of approximants to $1/3$ versus the inverse denominator $1/m$ (red squares, filled from one side and open from the other), its extrapolation (dotted line and black star, with an uncertainty smaller than the symbol), and the current at $\Delta=1/2$ itself (red line), with the double arrow marking the jump. (c)~The corresponding plot for angles $\eta$ that are not roots of unity, versus the distance from $\pi/3$ (red diamonds, filled and open for the two sides), with their extrapolation to zero distance (black star).}\end{figure*}

In the infinite chain, the current at a root of unity with low denominator is larger than the current at the surrounding values of the anisotropy. Across all roots of unity with denominator up to 200 and $\varepsilon=1$, points with $m\geq40$ trace out a continuous background of $J_\infty$, above which the small-denominator points stand out [Fig.~\ref{fig:jump}(a)]. At $\Delta=1/2$ and $\varepsilon=2$ the current takes the value $J_\infty(1/3)=\num{0.382542}$, while along four sequences of approximants, roots of unity $l/m$ whose denominators grow on their approach to $1/3$, two from each side and with denominators up to 4000, it tends to $\num{0.360768(7)}$ [Fig.~\ref{fig:jump}(b)]. We define the relative jump at $p/q$ as
\begin{equation}
D=1-\frac{\lim_{l/m\to p/q}J_\infty(l/m)}{J_\infty(p/q)} ,
\label{eq:D}
\end{equation}
and obtain $D=\num{0.05692(2)}$ at $1/3$ for $\varepsilon=2$ from the extrapolation along the approximants. At $\varepsilon=1$ the same procedure gives $D=\num{0.05909(1)}$ (Table~\ref{tab:S1} and Fig.~\ref{fig:S2}).

The limit does not depend on the path along which the angle approaches $\eta=\pi/3$. Angles that are not roots of unity, for which the transfer matrix does not truncate and the computation keeps up to $\num{1.2e6}$ levels, give the same lower value from both sides [Fig.~\ref{fig:jump}(c)], and their extrapolation to $\eta=\pi/3$ yields $D=\num{0.0569237}$. For such angles the current of the infinite chain is taken as $\zeta$ over the top eigenvalue of the matrix kept to that many levels, a value that stops changing as levels are added (Appendix~\ref{si:jumps}). There is no jump at an irrational target. Along the continued-fraction convergents of $l/m\to\sqrt2-1$ the current converges to the value computed directly at that angle, to a relative $\num{9.0e-9}$ at $2378/5741$ (Appendix~\ref{si:jumps}).

\subsection{Its origin in a phase that only the commensurate chain pins}

The jump arises because, around a root of unity, the transfer matrix is periodic up to a slowly drifting phase, and its leading eigenvalue is the largest permitted by any value of that phase. A slowly drifting phase of this kind also describes Bloch electrons in a magnetic field close to a rational flux, whose states locally resemble those of the rational system~\cite{wilkinson1987}. Consider an approximant whose angle $\eta=\pi l/m$ is close to the root-of-unity angle $\eta_0=\pi p/q$ with coprime integers $p$ and $q$, so that $\eta-\eta_0=\pi w/(mq)$ for a small integer $w=lq-pm$. The matrix entries in Eq.~\eqref{eq:T} depend on the level through $k\eta=k\eta_0+k(\eta-\eta_0)$. Across one block of $q$ levels, the first term returns to its initial value modulo $\pi$, and the second shifts by $\pi w/m$. Block by block, the Jacobi matrix of the approximant is therefore the period-$q$ matrix of the commensurate chain with a phase $\varphi=k(\eta-\eta_0)$ added to its arguments, and $\varphi$ drifts slightly from each block to the next [Fig.~\ref{fig:schematic}(b)]. Here $E_\mathrm{top}(\varphi)$ denotes the largest eigenvalue of a ring of $q$ sites with the site energies $d_r$ and the bonds $b_r$ given by
\begin{equation}
\begin{aligned}
d_r(\varphi)&=2\left|\cos(\eta_0 s-r\eta_0-\varphi)\right|^2,\\
b_r(\varphi)&=\left|\sin(2\eta_0 s-r\eta_0-\varphi)\right|\,\left|\sin[(r+1)\eta_0+\varphi]\right| ,
\end{aligned}
\label{eq:ring}
\end{equation}
where the index $r=0,\dots,q-1$ labels the sites of the ring and the bond $b_r$ couples site $r$ to site $r+1$ modulo $q$. Owing to the slow variation of the phase, the largest eigenvalue of the long Jacobi matrix tends to the largest band top contained in it (Appendix~\ref{app:bandtop}),
\begin{equation}
\lim_{l/m\to p/q}\tau_1(l/m)=\max_\varphi E_\mathrm{top}(\varphi).
\label{eq:limit}
\end{equation}
The commensurate chain corresponds to $\varphi=0$, at which the bond $b_{q-1}$ vanishes and the ring opens into the chain of $q$ levels, so that $\tau_1(p/q)=E_\mathrm{top}(0)$. Since $\zeta$ is continuous in $\eta$, the combination of Eqs.~\eqref{eq:Jinf},~\eqref{eq:D} and~\eqref{eq:limit} gives
\begin{equation}
D=1-\frac{E_\mathrm{top}(0)}{\max_\varphi E_\mathrm{top}(\varphi)} .
\label{eq:Dbt}
\end{equation}

For finite bath coupling, the relative jump $D$ in Eq.~\eqref{eq:Dbt} is strictly positive at each root of unity. Near $\varphi=0$ the severed bond reopens as $b_{q-1}\simeq c_0|\varphi|$ with the positive coefficient $c_0=|\sin(2\eta_0s-(q-1)\eta_0)|$, and first-order perturbation theory gives the shift $c|\varphi|$ in the top eigenvalue with the slope $c=2c_0v_0v_{q-1}$, where $v_0$ and $v_{q-1}$ are the end components of the positive eigenvector for the open chain. The band top thus kinks upward at $\varphi=0$, with its maximum lying elsewhere, and $D>0$. The period of the band top is $\pi/q$, since a shift of the phase by $\pi/q$ relabels the sites around the ring. It is also even in $\varphi$ and largest at $\varphi=\pi/(2q)$, because only the constant term of the characteristic polynomial of the ring depends on $\varphi$ (Appendix~\ref{app:bandtop}). The other entries of the ring also change linearly with $\varphi$, but because the band top is even their first-order contributions cancel, so $c|\varphi|$ is the whole first-order change and the two sides of the approach share the slope $c$. Evaluating Eq.~\eqref{eq:Dbt} gives $D=\num{0.0569237}$ for $1/3$ and $D=\num{0.0059719}$ for $2/7$ with $\varepsilon=2$, in agreement with the approximants [$\num{0.05692(2)}$ and $\num{0.005971(6)}$]. The size of the jump is set by how the band top depends on $\varphi$.

\subsection{Its dependence on the driving}
\label{sec:drive}

\begin{figure*}[t]
\includegraphics[width=\textwidth]{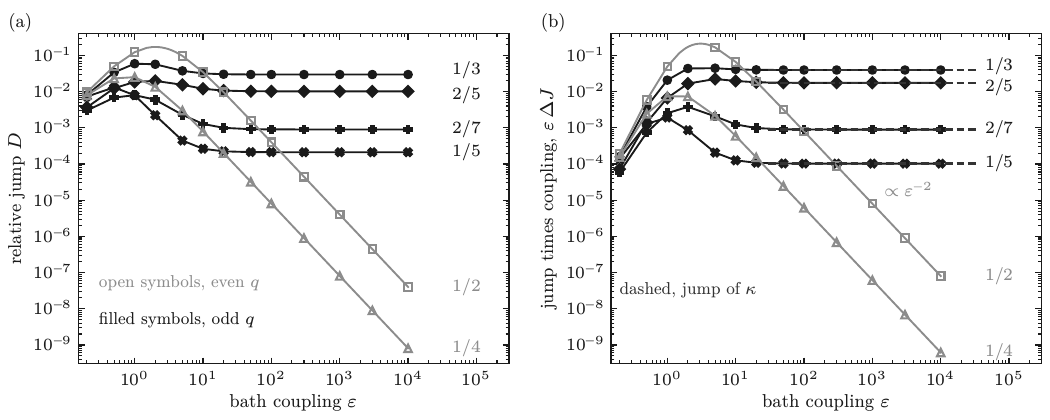}
\caption{\label{fig:drive}Dependence of the jump on the bath coupling $\varepsilon$. (a)~Relative jump $D$ of the infinite-chain current at six roots of unity $\eta=\pi p/q$, with filled black symbols for odd $q$ (circles $1/3$, diamonds $2/5$, plus signs $2/7$, crosses $1/5$) and open gray symbols for even $q$ (squares $1/2$, triangles $1/4$). (b)~The jump of the current multiplied by the coupling, $\varepsilon\,\Delta J$, with the same symbols. The dashed lines are the jump of the strong-driving coefficient $\kappa$ at each odd-$q$ point, labeled by $p/q$. For even $q$ the product falls as $\varepsilon^{-2}$. The uncertainties of the extrapolated values are smaller than the symbols. In both panels the curves are labeled by $p/q$ at their right end, and the lines joining the points guide the eye, except for $1/2$, whose line follows a dense scan of the coupling through $\varepsilon=2$.}
\end{figure*}

The jump grows largest when the bath coupling becomes comparable to the exchange. The relative jump $D$ from Eq.~\eqref{eq:Dbt} is shown in Fig.~\ref{fig:drive}(a) at six roots of unity for couplings from $0.2$ to $10^4$. For $\Delta=0$, with the chain free and $p/q=1/2$, the jump reaches $D=\num{0.123}$ at $\varepsilon=1$ and is largest as $\varepsilon\to2$, where the parameter $s$ is singular but the current is not. There $J_\infty(1/2)\to1/2$, the limit beside it tends to $\sqrt2-1$, and $D\to3-2\sqrt2=\num{0.1716}$, closed forms that the computed values match to $\num{6e-10}$. This is the jump at $q=2$. Reference~\cite{manna2024} reported that the largest current reached by varying the coupling jumps at $\Delta=0$ and that finite chains smooth this jump. For $1/4$ it amounts to $\num{0.0251}$ at $\varepsilon=1$ (Table~\ref{tab:S1}). Direct extrapolation of approximants reproduces the limiting current of Eq.~\eqref{eq:limit} at $\varepsilon=1$, 10, 100, 1000 and $10^4$, with the largest difference amounting to $\num{7.0e-7}$.

At strong driving the roots of unity separate by the parity of their denominator. For odd $q$ the jump of the current, $\Delta J=J_\infty(p/q)-\lim J_\infty(l/m)$, falls as $1/\varepsilon$ (a $\Delta$ in front of the current or of $\kappa$ denotes a jump and not the anisotropy), and $\varepsilon\,\Delta J$ tends to a constant. For even $q$ it falls as $\varepsilon^{-3}$ [Fig.~\ref{fig:drive}(b)]. The constants for odd $q$ are the jumps of the strong-driving coefficient defined in Sec.~\ref{sec:strong}. At $\varepsilon=10^4$ the ratio of $\varepsilon\,\Delta J$ to that jump differs from one by at most $\num{4.1e-7}$ at $1/3$, $2/5$, $2/7$ and $1/5$. The exponent of $\varepsilon\,\Delta J$ for $\varepsilon\geq10^3$ is $\num{-2.000}$ at $1/2$ and at $1/4$.

\section{The length that resolves the jump}
\label{sec:length}

\begin{figure*}[t]
\includegraphics[width=\textwidth]{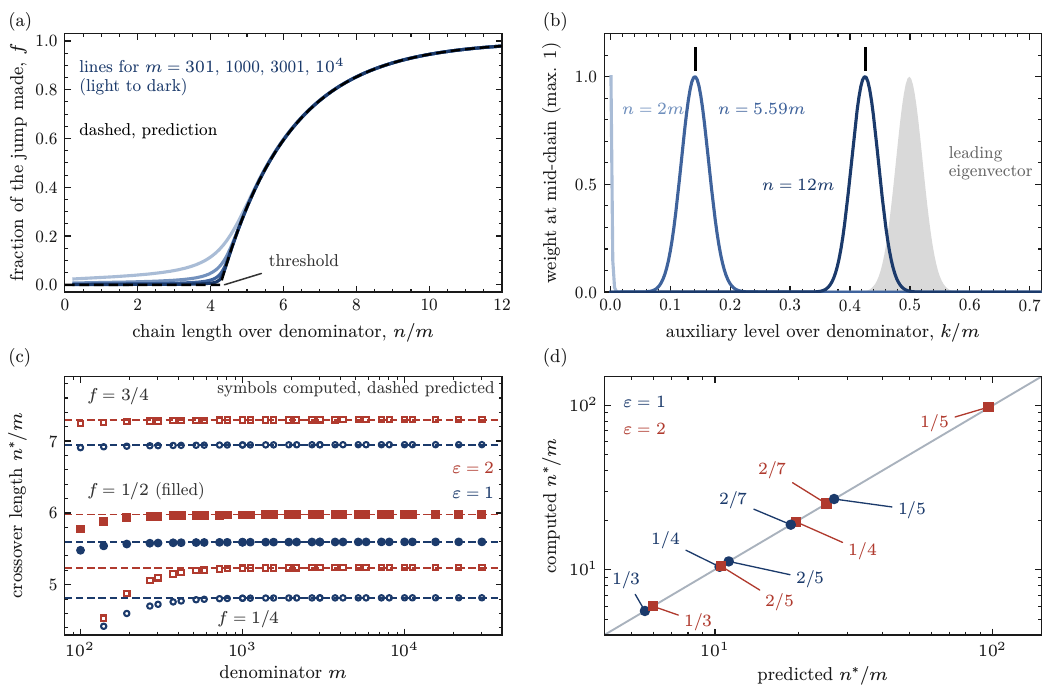}
\caption{\label{fig:crossover}A chain resolves the jump at a length set by a tunneling time. (a)~Fraction $f$ of the jump made on $n$ sites, Eq.~\eqref{eq:frac}, against $n/m$ for approximants $k/(3k+1)$ of $1/3$ with denominators $m=301$, 1000, 3001 and $10^4$ at $\varepsilon=1$ (blue lines, light to dark). The dashed line is Eq.~\eqref{eq:saddle}, with its threshold marked. (b)~Distribution over the auxiliary levels $k$ of the walks that make up $Z_n$, at the middle of the chain, for $m=1000$ and three chain lengths $n$ (blue lines, labeled). Each curve is summed over blocks of three levels and normalized to its maximum. Gray fill shows the squared leading eigenvector of the transfer matrix. Black ticks mark the turning points that Eq.~\eqref{eq:saddle} gives for the two longer chains. (c)~Crossover length $n^*_f$ over $m$ against $m$ for the same sequence, with blue circles for $\varepsilon=1$ and red squares for $\varepsilon=2$. Filled symbols are $f=1/2$ and open symbols $f=1/4$ and $f=3/4$, the latter at the larger lengths. Dashed lines show Eq.~\eqref{eq:nstar}. (d)~Crossover length $n^*/m$, with $n^*=n^*_{1/2}$, computed at $m\approx3000$ against Eq.~\eqref{eq:nstar}, for five roots of unity (labeled $p/q$) and two couplings. Each symbol is the mean of the two sides of the approach, which a vertical bar joins where it is longer than the symbol. The gray line marks equality.}
\end{figure*}

\subsection{A crossover length set by the detuning}

A finite chain cannot distinguish a root of unity from its neighbors until it is sufficiently long. On $n$ sites the current at an approximant $l/m$ of $p/q$ starts at the commensurate-chain current $J_{p/q}(n)$ and crosses over to its own limit $J_\infty(l/m)$. This crossover is followed through the fraction of the jump already made,
\begin{equation}
f(n)=\frac{J_{p/q}(n)-J_{l/m}(n)}{J_\infty(p/q)-J_\infty(l/m)},
\label{eq:frac}
\end{equation}
where $J_{l/m}(n)$ is the current of the chain of $n$ sites at the approximant, and the crossover length $n^*_f$ is defined as the length at which $f(n)$ first reaches the value $f$, with $n^*$ written for $n^*_{1/2}$. For four approximants of $1/3$ with denominators from 301 to $10^4$, our curves collapse onto one function of $n/m$ once $n\gtrsim5m$ [Fig.~\ref{fig:crossover}(a)]. Below this length the fraction is small and falls as $1/m$. At $n=2m$ the product $mf$ lies between $\num{12.0}$ and $\num{12.3}$ for these four approximants, which lie below $1/3$.

In such sequences the crossover length is proportional to the denominator. The detuning from the root of unity, defined as the anisotropy-angle difference $\delta=\eta-\eta_0$, is $\pm\pi/(mq)$. The ratio $n^*_f/m$ stays flat across two decades of $m$ [Fig.~\ref{fig:crossover}(c)], and a power law $n^*_f\propto m^\alpha$ with exponent $\alpha$ fitted in four denominator windows yields values that converge to one (Appendix~\ref{si:crossover}). In the window $3000\leq m\leq30000$ twelve fits, done for $\varepsilon=1$ and 2, on both sides of $1/3$ and for $f=1/4$, $1/2$ and $3/4$, give exponents from $\num{0.9994}$ to $\num{1.0002}$ with mid-value and half-range $\alpha=\num{0.9998(4)}$, which describe the spread among fits, not a statistical uncertainty. Iteration of the symmetrized transfer matrix in logarithmic variables instead gives $\alpha=\num{0.9994(12)}$ in the window $1000\leq m\leq10000$. Above $m=10^4$ the lengths come from the iteration with stored exponents (Appendix~\ref{si:crossover}), and the formula derived next reproduces them without iterating. For an anisotropy that is not a root of unity, there is no denominator, and the crossover length is inversely proportional to the detuning.

\subsection{The crossover as a tunneling time}
\label{sec:tunnel}

This linear law and its coefficient follow from the decay of the eigenvectors of the transfer matrix down to level 0. With the detuning $\delta$ the phase of Sec.~\ref{sec:finite} is $\varphi_k=k\delta$. An eigenvector of the Jacobi matrix whose eigenvalue $E$ lies between $E_\mathrm{top}(0)$ and $\max_\varphi E_\mathrm{top}$ is confined to the levels where $E_\mathrm{top}(\varphi_k)\geq E$, decaying outside them at a rate $\gamma(\varphi;E)$ per level that is fixed at each phase by the ring of Eq.~\eqref{eq:ring} (Appendix~\ref{app:instanton}), in the spirit of the exponentially weighted (Agmon-type) estimates for Harper's operator near a rational flux, whose spectral problem reduces to a $q\times q$ system~\cite{helffer1990}. The weight on level 0 of such an eigenvector $\psi_E$, normalized to one, is therefore
\begin{equation}
\begin{aligned}
\psi_E(0)^2&\sim\exp\left[-\frac{2W(E)}{|\delta|}\right],\\
W(E)&=\int_0^{\varphi_t(E)}\gamma(\varphi;E)\,d\varphi ,
\end{aligned}
\label{eq:action_A}
\end{equation}
where the exponent $W(E)$ is the decay rate integrated from level 0 to the turning point $\varphi_t(E)$, at which $E_\mathrm{top}(\varphi_t)=E$. Since $Z_n=\sum_j\psi_j(0)^2E_j^n$, a sum over the eigenvalues $E_j$ and the normalized eigenvectors $\psi_j$, for small $|\delta|$ this sum is dominated by the eigenvalue that maximizes $n\ln E-2W(E)/|\delta|$. The current of the chain of $n$ sites follows parametrically in $E$,
\begin{equation}
n|\delta|=\Theta(E)\equiv2E\,\frac{dW}{dE},\qquad J(n)=\frac{\zeta}{E}.
\label{eq:saddle}
\end{equation}
The function $\Theta(E)$ so defined is a tunneling time. In the picture of walks on the auxiliary levels, $\Theta(E)/|\delta|$ is the number of steps a walk spends tunneling from level 0 to the turning point and back. This time increases with $E$ from a finite threshold $\Theta_c$,
\begin{equation}
\Theta_c=\lim_{E\to E_\mathrm{top}(0)}\Theta(E)=\frac{\pi E_\mathrm{top}(0)}{qc},
\label{eq:tauc}
\end{equation}
where $c$ is the slope of the kink of Sec.~\ref{sec:finite}, and diverges as $E$ approaches the maximum of the band top. For $n|\delta|<\Theta_c$ no eigenvalue above $E_\mathrm{top}(0)$ competes with the states that live at level 0, and the chain carries the commensurate current. Beyond the threshold the current falls continuously toward the limit $J_\infty(l/m)$ of the approximants. The fraction $f$ of the jump is made at the eigenvalue $E_f$ and the length $n^*_f$ given by
\begin{equation}
n^*_f=\frac{\Theta(E_f)}{|\delta|},\qquad \frac{1}{E_f}=\frac{1-f}{E_\mathrm{top}(0)}+\frac{f}{\max_\varphi E_\mathrm{top}(\varphi)} ,
\label{eq:nstar}
\end{equation}
which for approximants with $|lq-pm|=1$, where $|\delta|=\pi/(mq)$, reads $n^*_f/m=q\Theta(E_f)/\pi$. No parameter in them is adjusted to the computed lengths. Appendix~\ref{app:instanton} gives the sensitivity of the prediction to the ring that enters it, and Table~\ref{tab:S5} compares it with four cases outside Table~\ref{tab:nstar}.

Equation~\eqref{eq:nstar} reproduces the computed crossover lengths without a fitted parameter. At $1/3$ and $\varepsilon=1$, Eq.~\eqref{eq:nstar} gives $n^*_f/m=\num{4.817}$, $\num{5.596}$ and $\num{6.950}$ for $f=1/4$, $1/2$ and $3/4$, and at $\varepsilon=2$ it gives $\num{5.238}$, $\num{5.978}$ and $\num{7.299}$ (Tables~\ref{tab:nstar} and~\ref{tab:S3}). The runs at $m\approx30000$ agree with these six numbers on both sides of $1/3$, with a largest relative difference of $\num{1.7e-4}$ [Fig.~\ref{fig:crossover}(c)]. The full curve $f(n/m)$ is described by Eq.~\eqref{eq:saddle}, including the threshold at $n/m=q\Theta_c/\pi=\num{4.271}$, which the computed curves approach as $m$ grows. For $m=10^4$ the largest deviation above the threshold is $\num{0.015}$, occurring next to it [Fig.~\ref{fig:crossover}(a)]. For five roots of unity and two couplings, Eq.~\eqref{eq:nstar} is compared with the lengths computed at $m\approx3000$ in Table~\ref{tab:nstar}. For nine of the ten cases, the two sides of the approach agree with each other to \qty{2}{\percent}, their mean matching the prediction to \qty{0.2}{\percent}. In the tenth, $1/5$ at $\varepsilon=2$, the jump is \qty{0.22}{\percent} of the current, the two sides still differ by \qty{8}{\percent} at $m\approx3000$, and the prediction lies between them, \qty{1.1}{\percent} from their mean [Fig.~\ref{fig:crossover}(d)].

The walks of the auxiliary index that make up $Z_n$ show the tunneling directly. Their distribution across auxiliary levels in the middle of the chain, $P(k)=\langle0|T^{n-j}|k\rangle\langle k|T^{j}|0\rangle/Z_n$ with $j$ the integer part of $n/2$, does not depend on the normalization of the levels. For $m=1000$ at $\varepsilon=1$ and $n=2m$, \qty{95}{\percent} of this distribution lies below the level $0.004m$. At $n=n^*=5.59m$, the distribution forms a peak centered around $k=0.141m$, where Eq.~\eqref{eq:saddle} puts the turning point at $0.142m$, and at $n=12m$ the peak has moved to $0.425m$ (turning point $0.426m$), next to the leading eigenvector, which is centered around $m/2$ [Fig.~\ref{fig:crossover}(b)]. A relative change $\num{1e-6}$ of the couplings in the center of the leading eigenvector leaves $J(n)$ unchanged to all digits for $n\leq6m$, whereas the same change applied to level 10 shifts this quantity by $\num{1.1e-7}$. The crossover is therefore a gradual shift of the dominant walks to higher levels as the chain grows, and the leading eigenvector takes over only in much longer chains.

The law is linear in the denominator at each root of unity and coupling, but its coefficient $n^*/m$ varies widely with the root of unity and with the coupling (Table~\ref{tab:nstar}). It ranges from $\num{5.6}$ at $1/3$ to $\num{26.9}$ at $1/5$ for $\varepsilon=1$, and to 97 at $1/5$ for $\varepsilon=2$. It is larger where the jump is smaller, without being a function of the jump alone. At $\varepsilon=1$ the jumps at $2/7$ and $1/5$ are \qty{0.80}{\percent} and \qty{0.79}{\percent} of the current, and the coefficients are $\num{18.8}$ and $\num{26.9}$.

The saddle point of Eq.~\eqref{eq:saddle} assumes that $\Theta(E)$ increases with $E$, as it does in the ten cases of Table~\ref{tab:nstar}. At the odd denominators we computed this fails at strong driving, above a coupling $\varepsilon_\mathrm{tri}$ at which $\Theta$ stops increasing at $E_\mathrm{top}(0)$ (Appendix~\ref{si:crossover}), of $\num{4.830}$ at $1/3$, $\num{2.604}$ at $1/5$, $\num{11.90}$ at $2/5$ and $\num{12.52}$ at $2/7$, whereas at $1/4$ it holds up to $\varepsilon=50$. Above $\varepsilon_\mathrm{tri}$ the maximum of $n|\delta|\ln E-2W(E)$ jumps from $E_\mathrm{top}(0)$ to an interior energy where $n|\delta|$ reaches $\min_E2W(E)/\ln[E/E_\mathrm{top}(0)]$, and the current of the finite chain drops there in a step instead of crossing over gradually. We call this value of $n|\delta|$, at which the two maxima are equal, the Maxwell length. At $1/3$ the step carries about $\num{0.8}$ of the jump at $\varepsilon=10$ and $\num{0.9}$ at $\varepsilon=20$ and lies at $n/m=\num{15.92}$ and $\num{20.35}$, where Eq.~\eqref{eq:nstar} would give $n^*/m=\num{14.97}$ and $\num{17.53}$.

\begin{figure}[tb]
\includegraphics[width=\columnwidth]{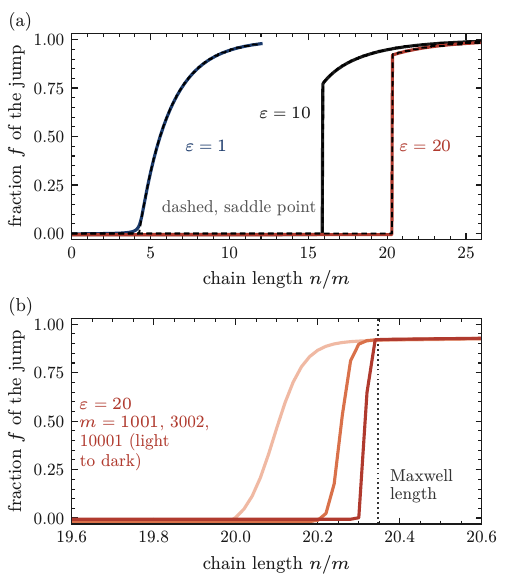}
\caption{\label{fig:step}At strong driving the crossover becomes a step. (a)~Fraction $f$ of the jump made by a chain of $n$ sites against $n/m$, for the approximant of $1/3$ with $m=10001$ at $\varepsilon=10$ and 20 and with $m=3002$ at $\varepsilon=1$. Dashed black lines show the saddle point of Eq.~\eqref{eq:saddle} taken as the global maximum of $n|\delta|\ln E-2W(E)$. (b)~The step at $\varepsilon=20$ for $m=1001$, 3002 and 10001, light to dark, with the Maxwell length dotted.}
\end{figure}

Finite chains computed from the transfer matrix of Eq.~\eqref{eq:T}, which we call exact chains, show the step. At $\varepsilon=20$ the current makes a tenth to nine tenths of the jump within $\num{0.21}$, $\num{0.069}$ and $\num{0.021}$ in $n/m$ for $m=1001$, 3002 and $10001$, a width that falls as $1/m$, around $n/m=\num{20.32}$ at the largest $m$, whereas at $\varepsilon=1$ the same range spans $\num{4.32}$. At $\varepsilon=10$ the chain with $m=10001$ makes half the jump at $n/m=\num{15.90}$. At $1/5$, where $\varepsilon_\mathrm{tri}=\num{2.604}$, exact chains averaged over the two signs of $\delta$ show a step that carries part of the jump, the rest being made gradually. For $m=30001$ and $30004$ its steepest rise lies at $n/m=\num{526.9}$ for $\varepsilon=5$ and $\num{1146.2}$ for $\varepsilon=10$, against Maxwell lengths of $\num{527.4}$ and $\num{1147.3}$, and its largest slope is $\num{3.39}$ and $\num{3.01}$ times that for $m=10001$ and $10004$. Equation~\eqref{eq:nstar} therefore applies below $\varepsilon_\mathrm{tri}$, where at the couplings computed, $\varepsilon=1$ and 2 at $1/3$, the threshold function increases and the maximum stays at $E_\mathrm{top}(0)$. Just above it the dip of $\Theta$ is small, its minimum being $\num{0.9994}$ of the threshold for $1/3$ at $\varepsilon=5$, where the formula still describes the computed lengths (Table~\ref{tab:S5}), while at $\varepsilon=10$ and 20 it fails as given above. The step and its sharpening with $m$ are shown in Fig.~\ref{fig:step}.

\begin{table}[tb]
\caption{\label{tab:nstar}Crossover length $n^*$, the number of sites at which a chain at an approximant $l/m$ has made half of the jump, in units of the denominator $m$, at five roots of unity $\eta=\pi p/q$ and two bath couplings $\varepsilon$. The column $D$ is the relative jump of the infinite-chain current, Eq.~\eqref{eq:Dbt}. The computed column gives $n^*/m$ for the approximant with $|lq-pm|=1$ and $m\approx3000$ on each side of $p/q$. The predicted column is Eq.~\eqref{eq:nstar} with $f=1/2$. The last column is $q\Theta_c/\pi$, from Eq.~\eqref{eq:tauc}, the value of $n/m$ below which the chain carries the commensurate current.}
\begin{ruledtabular}
\begin{tabular}{llS[table-format=1.2]S[table-format=3.3]S[table-format=3.3]S[table-format=3.3]S[table-format=2.3]}
 & & & \multicolumn{2}{c}{$n^*/m$ computed} & {$n^*/m$} & \\
\cline{4-5}
$p/q$ & $\varepsilon$ & {$D$ (\%)} & {$l/m>p/q$} & {$l/m<p/q$} & {predicted} & {$q\Theta_c/\pi$}\\
\hline
1/3 & 1 & 5.91 & 5.602 & 5.595 & 5.596 & 4.271\\
1/3 & 2 & 5.69 & 5.988 & 5.976 & 5.978 & 4.735\\
1/4 & 1 & 2.51 & 10.421 & 10.381 & 10.392 & 8.521\\
1/4 & 2 & 1.35 & 19.679 & 19.461 & 19.542 & 17.166\\
2/5 & 1 & 1.85 & 11.227 & 11.209 & 11.210 & 8.540\\
2/5 & 2 & 1.98 & 10.528 & 10.497 & 10.504 & 8.005\\
2/7 & 1 & 0.80 & 18.845 & 18.704 & 18.753 & 14.304\\
2/7 & 2 & 0.60 & 25.449 & 24.950 & 25.166 & 19.328\\
1/5 & 1 & 0.79 & 27.102 & 26.713 & 26.857 & 23.724\\
1/5 & 2 & 0.22 & 101.575 & 93.709 & 96.594 & 88.966\\
\end{tabular}
\end{ruledtabular}
\end{table}

\subsection{What a finite chain shows}
\label{sec:finitechain}

\begin{figure}[tb]
\includegraphics[width=\columnwidth]{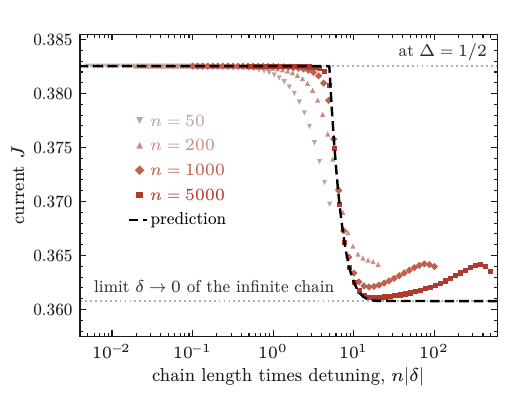}
\caption{\label{fig:profile}What a finite chain shows around $\Delta=1/2$ at $\varepsilon=2$. Current of chains with $n=50$, 200, 1000 and 5000 sites at anisotropy angles $\pi/3\pm\delta$ (mean of the two signs of $\delta$) versus $n|\delta|$, symbols keyed in the panel. The dashed line is Eq.~\eqref{eq:saddle}, dotted lines the infinite-chain current at $\Delta=1/2$ and, below it, the limit for $\delta\to0$. For large $n|\delta|$ the data rise above the prediction, where higher-order terms in $\delta$ come in.}
\end{figure}

On a finite chain the jump appears as a plateau of finite width. By Eq.~\eqref{eq:saddle} the current of a chain at detuning $\delta$ depends on $n$ and $\delta$ through the product $n|\delta|$, apart from the smooth dependence of the current on the anisotropy. It keeps the commensurate value for $n|\delta|<\Theta_c$ and falls to the lower branch beyond. The current of chains of 50 to 5000 sites at $\varepsilon=2$ and angles $\eta=\pi/3\pm\delta$, averaged over the two signs of $\delta$ to remove the part of the smooth dependence that is odd in $\delta$, is shown in Fig.~\ref{fig:profile}. At $\delta=0$ the current equals $J_\infty(1/3)$ to $\num{1.1e-16}$, the rounding error of the arithmetic, for all four lengths. The chains of 1000 and 5000 sites follow Eq.~\eqref{eq:saddle}, for which $\Theta_c=\num{4.96}$ at this root of unity and coupling. For $n|\delta|\leq30$ they deviate from it by at most $\num{3.2e-3}$ and $\num{1.3e-3}$ on the grid of Fig.~\ref{fig:profile}, while at the threshold itself their averaged current lies \qty{0.91}{\percent} and \qty{0.43}{\percent} below the plateau, a deviation that falls as the chain grows. The chains of 50 and 200 sites leave the plateau earlier, and at large $n|\delta|$ all four rise above the lower branch, the limit of the current along neighboring anisotropies, where terms of higher order in $\delta$ enter. A chain of $n$ sites therefore resolves the jump at $\Delta=1/2$ only at detunings $|\delta|$ above about $5/n$, this threshold being $0.1$ for 50 sites and $\num{1e-3}$ for 5000. The discontinuity belongs to the order of limits in which the length is taken to infinity first. What a finite chain can show is this scaling form. At any length the current near a root of unity is a function of $n|\delta|$ with a plateau and a drop, and the jump of the infinite chain is its limit.

In real space the crossover shows as magnetized layers at the two ends of the chain. At $1/3$ and $\varepsilon=1$ the commensurate chain carries no magnetization. Away from its end sites, the expectation value of the Pauli matrix $\sigma^\mathrm{z}$ at a site stays below $10^{-12}$. For the approximant $101/302$, in chains of 853, 1707 and 3413 sites, the magnetization a tenth of the chain from the left end is $\num{0.005}$ at $n=n^*/2$, $\num{0.077}$ at $n^*$ and $\num{0.126}$ at $2n^*$, with the opposite sign at the right end. Appendix~\ref{si:magnetization} shows the profiles. At $4n^*$ the layers have receded toward the ends, the magnetization being $\num{0.091}$ a tenth and $\num{0.020}$ a quarter of the chain from the end. The lower current of the long approximant thus comes with spin piling up next to the baths.

\section{Strong driving}
\label{sec:strong}

\subsection{The coefficient of the current}
\label{sec:kappa}

At large coupling the baths freeze the end spins, and the current falls as $1/\varepsilon$. In an expansion of Eq.~\eqref{eq:T} for $\varepsilon\to\infty$ (Appendix~\ref{app:zeno}), level 0 decouples from the other levels, $\zeta$ tends to $4\sin^2\eta/\varepsilon$, and the levels $k=1,\dots,m-1$ carry the Jacobi matrix $M$, which has diagonal entries $M_{k,k}$ and off-diagonal entries $M_{k,k+1}$ given by
\begin{equation}
M_{k,k}=2\sin^2(k\eta),\qquad M_{k,k+1}=\left|\sin(k\eta)\sin[(k+1)\eta]\right| .
\label{eq:M}
\end{equation}
The strong-driving coefficient of the current is therefore
\begin{equation}
\kappa\equiv\lim_{\varepsilon\to\infty}\varepsilon J_\infty=\frac{4\sin^2\eta}{\lambda_{\max}(M)},
\label{eq:kappa}
\end{equation}
where $\lambda_{\max}(M)$ is the largest eigenvalue of $M$. The chain length goes to infinity before the coupling does. A chain of given length shows the jumps of $\kappa$ only beyond its crossover length, which at $1/3$ grows from $\num{5.6}$ denominators at $\varepsilon=1$ (Table~\ref{tab:nstar}) to the Maxwell length $\num{20.35}$ at $\varepsilon=20$. The corrections are of relative order $\varepsilon^{-2}$. Evaluated at finite coupling, $\varepsilon J_\infty$ approaches $\kappa$ with a residual falling as $\varepsilon^{-2}$ over three decades (Appendix~\ref{si:drive}).

The coefficient $\kappa$ and a known lower bound on the Drude weight are different functions of one and the same matrix. The comparison asks whether the coefficient of the strongly driven current follows the linear-response Drude weight across the anisotropies. The lower bound $D_{\mathrm{Z}}$ of Ref.~\cite{prosen2011open} on the high-temperature Drude weight is obtained at the same roots of unity from a transfer matrix. In the form in which we evaluate it, the bound is the reciprocal of the first diagonal element of the inverse of a matrix that equals $M/2$ up to a diagonal change of signs, and it reproduces the closed-form values of that reference (Appendix~\ref{si:bound}). The coefficient $\kappa$ instead uses the largest eigenvalue of $M$. The combination $2D_{\mathrm{Z}}$ equals $\kappa$ at $\Delta=0$ and at no other root of unity we computed. At $\Delta=1/2$, $\kappa=4/3$ is the coefficient of the decay at large coupling of the current found in Ref.~\cite{prosen2011exact}, while $2D_{\mathrm{Z}}=9/8$, and at all 1100 roots of unity with $m\leq60$ away from $\Delta=0$ we find $\kappa>2D_{\mathrm{Z}}$ [Fig.~\ref{fig:kappa}(a)]. The two quantities also differ in where they are discontinuous, $2D_{\mathrm{Z}}$ jumping at every root of unity computed, by $\num{0.375}$ at $1/3$ and $\num{0.167}$ at $1/4$, whereas $\kappa$ jumps at odd denominators only, by \qty{3.0}{\percent} of its value at $1/3$. The jumps of the bound of Ref.~\cite{prosen2011open} follow from its closed form $D_{\mathrm{Z}}=\sin^2(\pi l/m)\,m/[2(m-1)]$, as given in Ref.~\cite{prosen2013quasilocal}, which our values reproduce to $\num{1.6e-12}$ at all 12231 roots of unity with $m\leq200$. Approximants have $m\to\infty$, so $2D_{\mathrm{Z}}$ jumps by $\sin^2\eta_0/(q-1)$ at every root of unity, odd or even. The improved bound $D_{\mathrm{K}}=[\sin^2(\pi l/m)/\sin^2(\pi/m)][1-(m/2\pi)\sin(2\pi/m)]$ of the same reference~\cite{prosen2013quasilocal}, in the same convention and equal to $\kappa/2$ at $\Delta=0$, coincides with the high-temperature Drude weight obtained from the thermodynamic Bethe ansatz~\cite{urichuk2019}. The factor 2 makes the three quantities equal at $\Delta=0$. The bound $2D_{\mathrm{K}}$ also jumps at every root of unity, by $\num{0.173}$ at $1/3$ and $\num{0.060}$ at $1/4$, against $\num{0.040}$ and zero for $\kappa$, and of the 1100 roots of unity with $m\leq60$ away from $\Delta=0$, $\kappa$ exceeds $2D_{\mathrm{K}}$ at 720 and not at the other 380.

At weak coupling the bound enters the current itself. As $\varepsilon\to0$, level 0 decouples from the others, the product of the two entries of $T$ that join it to level 1 being $\varepsilon^2$ up to terms of order $\varepsilon^4$, and on the levels $k\geq1$ the symmetrized transfer matrix at $\varepsilon=0$ equals $2-M$ up to a diagonal change of signs. Second-order perturbation theory in the level 0 coupling then yields
\begin{equation}
J_\infty=\frac{\varepsilon}{2}-\frac{\varepsilon^3}{8D_{\mathrm{Z}}}
\label{eq:weak}
\end{equation}
up to terms of order $\varepsilon^5$ at every root of unity, where $1/(2D_{\mathrm{Z}})$ is the first diagonal element of $M^{-1}$. That the steady state at first order in the coupling yields the conserved charge from which $D_{\mathrm{Z}}$ follows is the construction of Ref.~\cite{prosen2011open}, and Eq.~\eqref{eq:weak} places the bound in the third-order term of the current. Since $1/D_{\mathrm{Z}}$ jumps by $2/(q\sin^2\eta_0)$, the relative jump at weak coupling is $D=\varepsilon^2/(2q\sin^2\eta_0)$ up to terms of order $\varepsilon^4$ for fixed odd or even denominator. The limit is not uniform in the denominator. For $p=1$, the coupling at which $D$ has fallen to half of this value decreases as $\num{7.1}\,q^{-2.06}$ for $q$ from 5 to 15, so the law holds only for couplings small compared with $1/q^2$. At all roots of unity with $m\leq30$ the coefficient of $\varepsilon^3$, extrapolated to $\varepsilon\to0$, equals $1/(8D_{\mathrm{Z}})$ to a relative $\num{7.3e-8}$, and at $\Delta=1/2$ the values of $D/\varepsilon^2$ from Eq.~\eqref{eq:Dbt} extrapolate to $\num{0.2222222222}$, against $2/9$. The same construction with $D_{\mathrm{K}}$ in place of $D_{\mathrm{Z}}$ misses by at least \qty{100}{\percent} at the six roots of unity of Fig.~\ref{fig:drive}. For $\varepsilon=0.05$, extrapolating approximants to $1/3$ reproduces the value of $D$ from Eq.~\eqref{eq:Dbt} to a relative $\num{4.1e-4}$ (Appendix~\ref{si:bound}).

\subsection{Why even denominators give no jump}
\label{sec:parity}

The jump of $\kappa$ is controlled by the strong-driving limit of the periodic matrix of Eq.~\eqref{eq:ring}, which describes a ring of $q$ sites with couplings that depend on the phase $\varphi$. This ring carries the site energies $2\sin^2(r\eta_0+\varphi)$ and the bonds $|\sin(r\eta_0+\varphi)\sin[(r+1)\eta_0+\varphi]|$, and, following the argument of Sec.~\ref{sec:finite}, the jump $\Delta\kappa$ of the coefficient is
\begin{equation}
\begin{aligned}
\Delta\kappa&\equiv\kappa(p/q)-\lim_{l/m\to p/q}\kappa(l/m)\\
&=4\sin^2\eta_0\left[\frac{1}{\mathcal{E}(0)}-\frac{1}{\max_\varphi\mathcal{E}(\varphi)}\right],
\end{aligned}
\label{eq:kjump}
\end{equation}
where $\mathcal{E}(\varphi)$ denotes the largest eigenvalue of that ring. This ring factorizes as a product, its matrix being $VCV$, where $V$ is the diagonal matrix of the $\sin(r\eta_0+\varphi)$ and $C$ is the circulant matrix with 2 on the diagonal and 1 between neighbors. In the basis that diagonalizes $C$, the momenta $2\pi j/q$ with $j=0,\dots,q-1$, the eigenvalues of $C$ are the weights $K_j=4\cos^2(\pi j/q)$, and $V^2$ couples the momentum $j$ only to $j\pm p$. The nonzero eigenvalues of $VCV$ are those of a second ring, whose sites are the momenta $j$. Its site energies are $K_j/2$, and it has a bond from each momentum $j$ to the momentum $j+p$, equal to $-\frac14e^{2i\varphi}$ times the geometric mean of the two weights (Appendix~\ref{app:dual}). The phase $\varphi$ enters only as a flux $2q\varphi$ through this ring.

\begin{figure*}[t]
\includegraphics[width=\textwidth]{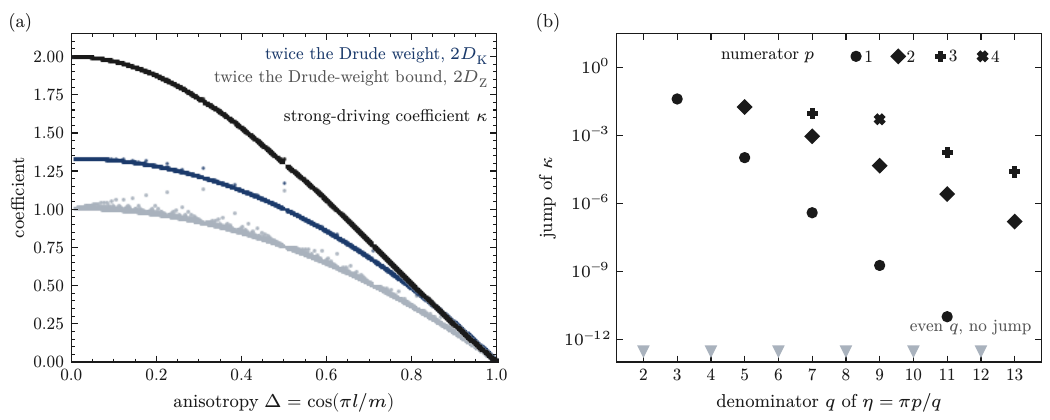}
\caption{\label{fig:kappa}The strong-driving coefficient. (a)~$\kappa$ (black dots), twice the lower bound $D_{\mathrm{Z}}$ on the Drude weight (gray dots) and twice the bound $D_{\mathrm{K}}$, which equals the Drude weight (blue dots), at every root of unity $\eta=\pi l/m$ with $m\leq200$, against the anisotropy. On this scale the jumps at low denominators are barely visible, and panel (b) gives those of $\kappa$. (b)~Jump of $\kappa$ at $\eta=\pi p/q$ against the denominator $q$. The marker gives the numerator $p$ (key in the panel), and for $p=4$ only $q=9$ was computed. Gray triangles on the lower axis correspond to even $q$ (where this jump is zero).}
\end{figure*}

For even $q$ the ring of momenta is cut, and its top eigenvalue no longer depends on the phase. The weight $K_j$ vanishes at $j=q/2$, the two bonds that touch that site vanish with it, and the flux can be removed by a change of the phases of the basis. Hence $\mathcal{E}(\varphi)$ does not depend on $\varphi$, and by Eq.~\eqref{eq:kjump} $\kappa$ does not jump. At the six even denominators we checked, $\mathcal{E}(\varphi)$ varies with $\varphi$ by at most $\num{3.8e-80}$, and setting the weight at $j=q/2$ to one by hand restores a variation between $\num{6.9e-5}$ and $\num{0.5}$. For odd $q$ no weight vanishes and the flux matters. For Hermitian matrices with given entry magnitudes, the top eigenvalue is largest when a change of the basis phases makes all off-diagonal entries positive. On a ring this is possible only if the product of the bond phases around it equals one, which here reads $(-1)^qe^{2iq\varphi}=1$. For odd $q$ that is $\varphi=\pi/(2q)$, and at $\varphi=0$ the flux is the least favorable one. Therefore $\max_\varphi\mathcal{E}=\mathcal{E}(\pi/2q)>\mathcal{E}(0)$, and at every odd denominator the coefficient jumps. At the 55 roots of unity where the phase was scanned, the maximum of the scan agrees with $\mathcal{E}(\pi/2q)$ to $\num{3.0e-64}$. At the larger ones the maximum is evaluated at $\pi/(2q)$. At every odd $q$ computed, the jumps are finite [Fig.~\ref{fig:kappa}(b)]. The jump of $\kappa$ at $3/7$ is $\num{8.742e-3}$, and those at the even denominators $3/10$ and $5/12$ are at most $\num{3.9e-81}$ in magnitude.

\subsection{A tunneling law for odd denominators}
\label{sec:tunnellaw}

\begin{figure*}[t]
\includegraphics[width=\textwidth]{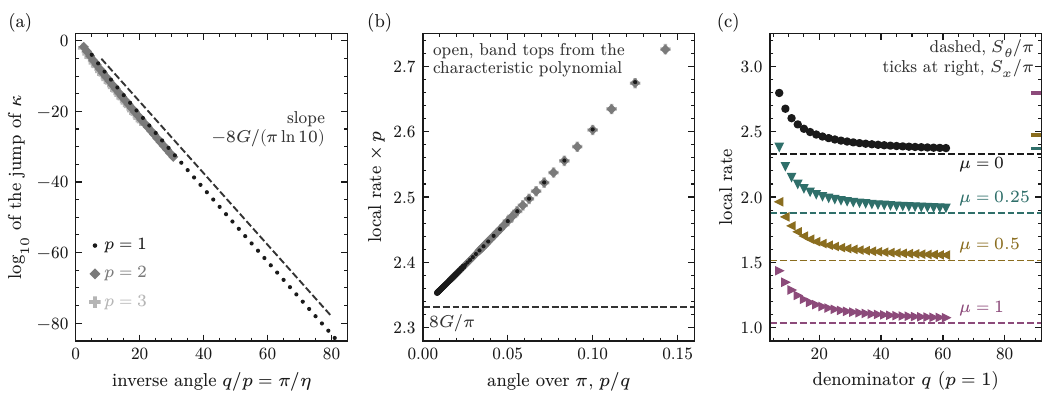}
\caption{\label{fig:tunnel}The jumps of the strong-driving coefficient follow a tunneling law. (a)~Decimal logarithm of the jump of $\kappa$ at $\eta=\pi p/q$ against $q/p$, for numerators $p=1$ (small black circles), 2 (gray diamonds) and 3 (light gray plus signs), drawn in increasing size so that coinciding points stay visible. The dashed line is the decay rate $8G/\pi$ of the natural logarithm per unit $q/p$, which is the slope $-8G/(\pi\ln10)$ on this decimal scale, with $G$ Catalan's constant. (b)~Local decay rate multiplied by $p$ against $p/q$, same symbols, approaching $8G/\pi$ as $p/q\to0$. Open symbols are band tops from the characteristic polynomial, up to $q/p=60$ and, for $p=1$, up to $q=121$, and the dashed line is $8G/\pi$. (c)~Local rates for $p=1$ with the momentum weights lifted by the parameter $\mu$ (labeled). Dashed lines show the action $S_\theta$ along the momentum $\theta$ divided by $\pi$, for each $\mu$, and ticks at the right edge the action $S_x$ along $x$ divided by $\pi$, in the color of its $\mu$.}
\end{figure*}

Jumps of $\kappa$ at odd denominators decay exponentially in the inverse of the anisotropy angle $\eta$ when $\eta$ is small. Since $\kappa$ takes the same value at $l/m$ and $(m-l)/m$, for $\Delta<0$ the angle is $\pi-\eta$. For $p=1$ the jumps take the values $\num{0.0395}$, $\num{1.020e-4}$, $\num{3.799e-7}$ and $\num{1.801e-9}$ at $q=3$, 5, 7 and 9, and the sequence goes on to fall by a nearly constant factor for each step of two in $q$, down to $\num{6.9e-85}$ at $q=81$ [Fig.~\ref{fig:tunnel}(a) and Table~\ref{tab:S4}]. The ratio of the jumps at $1/9$ and $1/5$ comes out as $\num{1.765456e-5}$ from the ring of Eq.~\eqref{eq:kjump} evaluated at 100 digits and as $\num{1.765453e-5}$ from extrapolation of approximants in double precision, and the difference of $\num{2.4e-11}$ lies inside the uncertainty $\num{1.0e-10}$ of the second. The decay is not set by the denominator alone. The jump at $2/9$ is $\num{0.4437}$ times the jump at $1/5$, where a suppression that depended on the denominator only would give the $\num{1.8e-5}$ found at $1/9$.

Semiclassically, a tunneling action gives the rate. With the coordinate $x=r\eta+\varphi$ and the momentum $\theta$ conjugate to it, a step of one site shifts $x$ by $\eta$, so $\eta$ plays the role of Planck's constant, and the ring $VCV$ has the classical symbol $4\sin^2x\cos^2(\theta/2)$, the function that the matrix becomes when the shift by one site is replaced by the phase factor $e^{i\theta}$. This symbol takes its maximum value of 4 at $x=\pi/2$ and $\theta=0$. The top eigenvalue can depend on the offset $\varphi$ only through paths that wind once around the momentum direction at that energy.

For $p>1$ the coordinate advances by $p\pi$ around the ring, which therefore holds $p$ equivalent maxima of the symbol, and the top $p$ eigenvalues of the ring form a cluster split only by tunneling, whereas the next eigenvalue lies $\num{0.545}$ below at $p=2$, $q=41$ and $\num{0.549}$ below at $p=3$, $q=61$. In the strong-driving limit the constant term of the characteristic polynomial of App.~\ref{app:bandtop} reduces to a cosine, $\det[E-H(\varphi)]=Q(E)+c\cos(2q\varphi)$, with $Q(E)$ a polynomial independent of $\varphi$ and $c=2\cdot4^{1-q}$. The top eigenvalue therefore moves between $\varphi=0$ and $\varphi=\pi/(2q)$ by $2c/Q'(E_\mathrm{top})$, with $Q'$ the derivative of $Q$ at the top eigenvalue for $\varphi=0$, times a factor of $\num{1.000}$, $\num{0.828}$ and $\num{0.764}$ for $p=1$, 2 and 3, the same at the two denominators computed for each, $q=21$ and 31, 41 and 61, and 61 and 91. The barrier factors, the tunneling amplitudes between neighboring maxima of the symbol, sit in $Q'(E_\mathrm{top})$, the product of the distances to the other eigenvalues. Divided by the product over the eigenvalues outside the cluster, $c$ decays per unit $q/p$ at the rates $\num{2.24}$, $\num{4.55}$ and $\num{6.86}$ for $p=1$, 2 and 3, about $p$ barrier factors, and the $p-1$ splittings of the cluster cancel all but one. Accordingly, $-\ln$ of the shift divided by $q/p$ is $\num{2.293}$ for $p=2$ at $q=61$ and $\num{2.330}$ for $p=3$ at $q=91$, approaching $8G/\pi$ from below. Likewise, the wells of Harper's operator interact by tunneling, with exponentially small coefficients~\cite{helffer1988harper}.

A path that winds around the momentum direction cannot remain real. Between $\theta=0$ and $\theta=2\pi$ it requires $\sin^2x=1/\cos^2(\theta/2)>1$. Its action $S_\theta$ is given by
\begin{equation}
S_\theta=\int_0^{2\pi}\operatorname{arccosh}\frac{1}{|\cos(\theta/2)|}\,d\theta=8G,
\label{eq:action}
\end{equation}
with Catalan's constant $G\approx\num{0.916}$, and the jump is expected to scale as $e^{-S_\theta/\eta}=e^{-8Gq/(\pi p)}$ times a power of $q$. In the semiclassical treatment of Harper's equation~\cite{wilkinson1984}, the tunneling action at the edge of its spectrum is an integral of the same value, $8G$~\cite{fuchs2018}. The two operators are related but not identical. The symbol of Harper's equation is a sum of two cosines where ours is a product, and its Planck constant is set by the flux, so the two decay laws are not the same function of their parameters.

The computed jumps follow this law. The local rate between consecutive odd denominators $q$ and $q'$ is defined as $[\ln\Delta\kappa(q)-\ln\Delta\kappa(q')]/(q'-q)$. Multiplied by $p$, the rates for $p=1$, 2 and 3 fall on one curve as functions of $q/p$ and decrease toward $8G/\pi=\num{2.33249}$ [Fig.~\ref{fig:tunnel}(b), Table~\ref{tab:S4}], and at the largest denominators computed from the ring at 100 digits still lie \qty{1.4}{\percent} ($p=1$, $q=81$), \qty{3.7}{\percent} ($p=2$, $q=61$) and \qty{5.6}{\percent} ($p=3$, $q=61$) above it, as a power-law prefactor requires.

Fits of $\ln\Delta\kappa=\ln A+a\ln q-\beta q$ with prefactor $A$, power $a$ and rate $\beta$, with and without a further term in $1/q$, give $p\beta$ within \qty{0.61}{\percent} of $8G/\pi$ in each window starting at $q\geq15$ (Appendix~\ref{si:strong}). These percentages are deviations of the fitted rates from $8G/\pi$, not uncertainties of the fits. Finding each band top as the largest root of the characteristic polynomial extends the series to $q=121$ for $p=1$ and to $q/p=60$ for $p=2$ and 3. The local rates keep decreasing, down to \qty{0.90}{\percent} above $8G/\pi$ at $q=121$, and between the two odd denominators bracketing $q/p=60$ the three numerators give $\num{2.37484}$, $\num{2.37483}$ and $\num{2.37483}$ for $p$ times the rate.

The fitted power $a$ stays between $-2.5$ and $-2.8$ in windows starting at $q\geq15$. The power $-5/2$ is the one expected for a single tunneling splitting. At a nondegenerate maximum of the symbol the splitting carries a factor $\eta^{1/2}$, as for Harper's equation~\cite{fuchs2018}, and the factor $4\sin^2\eta_0$ of Eq.~\eqref{eq:kjump} adds $q^{-2}$. With the power fixed at $-5/2$, a fit of two parameters gives $p\beta$ within \qty{0.07}{\percent}, \qty{0.20}{\percent} and \qty{0.30}{\percent} of $8G/\pi$ for $p=1$, 2 and 3 over windows starting at $q/p\geq15$, and within \qty{0.02}{\percent} when a term in $1/q$ is added. The prefactor depends on $p$, and in the form $\Delta\kappa\simeq A\,q^{-5/2}e^{-8Gq/(\pi p)}$ the fits with the term in $1/q$ give $A\approx450$, 266 and 239 for $p=1$, 2 and 3. For a given angle the jump thus depends on the numerator through $A\,p^{-5/2}$, and only the exponential rate is a function of $\eta$ alone. A single splitting would make the whole jump a function of $\eta$, and for $p>1$ the ring holds $p$ wells.

Deforming the ring tells apart two actions that coincide for it, the one for tunneling along the momentum and the one for tunneling along the coordinate. To this end, $K_j$ is replaced by $K_j+4\mu^2$ with a real parameter $\mu$, lifting the vanishing weight at the edge of the momentum ring. The deformation, which has no spin-chain counterpart, separates the action $S_\theta(\mu)$ along $\theta$ from the action $S_x(\mu)$ for tunneling along $x$ (Appendix~\ref{app:dual}), which coincide at $\mu=0$. For $\mu=0.25$, $0.5$ and 1 the predicted rates $S_\theta/\pi$ are $\num{1.8763}$, $\num{1.5147}$ and $\num{1.0358}$, and the rates fitted between $q=31$ and $q=61$ are $\num{1.8763}$, $\num{1.5147}$ and $\num{1.0357}$, whereas $S_x/\pi$ would be $\num{2.371}$, $\num{2.477}$ and $\num{2.799}$ [Fig.~\ref{fig:tunnel}(c) and Table~\ref{tab:S6}]. Under the deformation the decay remains exponential, with the rate set by the momentum direction, whereas the value $8G$ belongs to the undeformed ring, whose weight vanishes at the momentum $\pi$.

\section{Discussion}
\label{sec:discussion}

The size of the jump, the crossover length and the strong-driving law belong to a single semiclassical picture, in which the phase drifting along the auxiliary levels near a root of unity plays the role of a slow coordinate. Its value at the maximum of the band top fixes the current of the infinite chain and hence the jump. The decay of the eigenvectors from that maximum toward level 0 fixes how long a chain must be to feel it, Eq.~\eqref{eq:saddle}. At any finite coupling the cut bond of the commensurate ring reopens linearly in the phase, producing a kink in the band top and a jump in the current at every root of unity, and the slope of this kink also sets the threshold of Eq.~\eqref{eq:tauc}. As the coupling grows, the level carrying the kink decouples and leaves behind the smooth flux dependence of the strong-driving ring, which exists only for odd denominators and is exponentially small in $1/\eta$. Accordingly, the even-denominator jumps disappear as $\varepsilon^{-3}$, and the odd ones cross over to $\Delta\kappa/\varepsilon$ [Fig.~\ref{fig:drive}(b)].

That the discontinuity exists is implicit in the exact solution. The present paper adds the size of the jump, the finite-length theory and the strong-driving law. The reduction of the transfer matrix at roots of unity and the formula~\eqref{eq:Jinf} are taken from Ref.~\cite{prosen2015review}, ballistic transport in the easy-plane regime was proven in Ref.~\cite{prosen2011exact}, and a completely discontinuous dependence of the spin Drude weight and of asymptotic spin-current rates on the anisotropy is known~\cite{ilievski2017}. The lower bound on the Drude weight is a fractal function of the anisotropy~\cite{prosen2011open}. In our evaluation this bound jumps at every root of unity computed, as the current does at finite driving, and the strong-driving coefficient differs from both, and from the Drude weight, which is discontinuous at every anisotropy, in jumping at odd denominators only. A fractal Drude weight also occurs in the sine-Gordon model~\cite{sinegordon2023}, and this arithmetic structure is therefore not special to the XXZ chain. The discontinuous decay rate of finite-size corrections reported in Ref.~\cite{manna2024} concerns a counting-field deformation of the same transfer matrix. The crossover length derived here describes how the current approaches its limiting value in a chain near, and not at, a root of unity. In the closed chain the time needed to tell a rational anisotropy from the values around it is a separate question, tied to the low-frequency conductivity~\cite{agrawal2020}. The length $n^*$ refers to the steady state and to the detuning of the anisotropy, and hence the two scales are not directly comparable.

Testing this current jump requires chains of several hundred sites. From Eq.~\eqref{eq:tauc}, in a chain of $n$ sites at $\varepsilon=2$, the current at $\Delta=1/2$ is distinguished from that of the neighbors only for detunings $|\delta|$ larger than about $5/n$. In chains of a few tens of sites this is an angle of order one tenth, where smooth anisotropy dependence prevails (Fig.~\ref{fig:profile}), and a 1000-site chain would show a plateau of half-width $\num{5e-3}$ in $\eta$, past which the current is lower by \qty{5.69}{\percent}. At $\Delta=0$, the jump is about three times larger ($D\to\num{0.1716}$ as $\varepsilon\to2$), and the threshold lies lower, $n|\delta|=\num{3.15}$ for $\varepsilon=1.9$ and $2.1$. The driving in Eq.~\eqref{eq:lindblad} injects spin incoherently into one end and extracts it from the other. Circuits whose boundary qubits are reset to fixed states are driven in a similar way~\cite{popkov2025circuits}, but we know of no device that holds several hundred spins coherent under such driving. A direct simulation of the Lindblad equation would check our evaluation of the exact steady state independently of the matrix-product route. It would not test an approximation. Such a check needs the current averaged over $\pi/3\pm\delta$, which removes the smooth dependence on the anisotropy, and finite-chain values rather than the long-chain limits of Eq.~\eqref{eq:saddle}. Those limits give no drop below $n|\delta|=\num{4.96}$ and \qty{5.2}{\percent} at $n|\delta|=10$. For $n=200$ and 500 the exact steady state gives drops of \qty{1.82}{\percent} and \qty{1.24}{\percent} at $n|\delta|=\num{4.96}$, and of \qty{4.40}{\percent} and \qty{4.82}{\percent} at $n|\delta|=10$. A chain of 5000 sites gives \qty{0.43}{\percent} and \qty{5.19}{\percent}. For the other roots of unity of Table~\ref{tab:nstar} the threshold lies higher and the plateau is lower.

The results hold for maximal polarization of the baths, where the matrix-product solution truncates at roots of unity. Partial polarization, dephasing and integrability breaking lie outside them, and so do the steady states, at strong dissipation, of chains whose boundary spins are polarized in arbitrary directions~\cite{popkov2020dirichlet,popkov2020inhom} and the steady state of a chain driven dissipatively at one end with a coherent boundary field at the other~\cite{oneend2026}, to which the analysis could be extended. The chain with an integrable impurity also has a matrix-product steady state, and the impurity leaves its long-chain current unchanged~\cite{matsui2024exact}, so the jumps may persist there. Away from maximal polarization the steady state is no longer a matrix product with the transfer matrix used here, and its auxiliary space does not truncate at roots of unity. Whether a jump survives there is not fixed by the exact solution and needs steady states computed without it. Equation~\eqref{eq:limit} rests on the argument of Appendix~\ref{app:bandtop}, and Eqs.~\eqref{eq:saddle}--\eqref{eq:nstar} on a saddle-point evaluation to exponential accuracy (Appendix~\ref{app:instanton}). Both are confirmed numerically and neither is proven with error bounds. The exponent $8G/\eta$ is asymptotic, and the local rates are still approaching it over the computed denominators, and neither the power of $q$ nor the dependence of the prefactor on $p$ is derived. For finite driving Eq.~\eqref{eq:Dbt} yields the jump at any root of unity, but we have not obtained its law for large denominators. Equation~\eqref{eq:weak} gives the jump only for couplings below a scale that falls as $q^{-2}$. For $p=1$ the jumps in $\kappa$ at $q\geq7$ do not exceed $\num{3.8e-7}$ (Table~\ref{tab:S4}). Their interest lies in the underlying mechanism, which is a tunneling of the same kind as the one that sets the crossover length at finite driving.

\section{Conclusions}
\label{sec:conclusions}

In the maximally driven XXZ chain, the current discontinuity exists only when the length is taken to infinity first. The jump size, the crossover length, and the strong-driving law specify what a chain of given length retains of it. In the infinite chain, the limit along the neighbors of a root of unity lies below the current at the root by an amount fixed by the phase dependence of a single eigenvalue (\qty{5.69}{\percent} of that current at $\Delta=1/2$ for $\varepsilon=2$). In a chain of $n$ sites, the jump gives way to a plateau of half-width about $5/n$ in the anisotropy angle at this root of unity, and the tunneling time of Eq.~\eqref{eq:saddle} gives the length at which the current leaves the plateau while magnetized layers form at the two ends. This length, which for the closest approximants is $\num{5.6}$ denominators at $1/3$ and $\varepsilon=1$ and 97 at $1/5$ and $\varepsilon=2$, marks a gradual crossover below the coupling $\varepsilon_\mathrm{tri}$ and a step at odd denominators above that coupling. Under strong driving, the plateau survives at odd denominators only, with its height falling as $e^{-8G/\eta}$ for small angles. Simulations of the Lindblad equation for a few hundred sites can check the finite-chain currents around $\Delta=1/2$ from the exact solution.

\emph{Data availability.} The code that produced the results and the calculation records of the numbers quoted in the text and plotted in the figures are deposited at Zenodo, \href{https://doi.org/10.5281/zenodo.24548540}{10.5281/zenodo.24548540}.

\begin{acknowledgments}
The author gratefully acknowledges the Department of Physics, Indian Institute
of Technology Delhi, for computational resources and support.
The author used Claude (Anthropic) to cross-check numerical values, equations, cross-references and citations against the calculation scripts and primary sources, and verified all of its output.
\end{acknowledgments}

\appendix

\section{Strong-driving limit of the transfer matrix}
\label{app:zeno}

For $\varepsilon>2\sin\eta$, the condition $\tan(\eta s)=-iX$, with the ratio $X=\varepsilon/(2\sin\eta)$, is solved by $\eta s=-\pi/2+iy$ for real $y$ fixed by $\coth y=-X$, so $y\simeq-1/X$ for large $X$. Substituting this solution into Eq.~\eqref{eq:T} gives the entries $T_{k,k}$ along the diagonal and $T_{k,k+1}$ above it as
\begin{equation}
\begin{aligned}
T_{k,k}&=2\sin^2(k\eta)+2\sinh^2y,\\
T_{k,k+1}&=\sin^2(k\eta)+\sinh^2(2y),
\end{aligned}
\label{eq:Tlarge}
\end{equation}
while the entries $T_{k+1,k}=\sin^2[(k+1)\eta]$ below the diagonal do not depend on $\varepsilon$. For $k\geq1$ these symmetrized entries coincide with those of Eq.~\eqref{eq:M} up to terms of order $X^{-2}$. Level 0, carrying the diagonal entry $2\sinh^2y$, is coupled to level 1 by $|\sinh(2y)\sin\eta|\simeq4\sin^2\eta/\varepsilon$, and this level therefore decouples as $\varepsilon\to\infty$, shifting the leading eigenvalue only at order $\varepsilon^{-2}$. Finally $\zeta=\varepsilon\sinh^2y\simeq4\sin^2\eta/\varepsilon$, which together with Eq.~\eqref{eq:Jinf} gives Eq.~\eqref{eq:kappa}.

\section{Leading eigenvalue of a slowly modulated periodic Jacobi matrix}
\label{app:bandtop}

Consider a symmetric tridiagonal matrix of dimension $N$ whose entries at level $k$ are those of a period-$q$ ring at the phase $\varphi_k$, with non-negative off-diagonal entries and with $\varphi_k$ changing by an amount of order $1/N$ per level. Two bounds give Eq.~\eqref{eq:limit}.

To obtain the lower bound, take the positive eigenvector of the ring at the maximizing phase $\varphi^*$, extend it periodically, and multiply the extended vector by an envelope of width $\Lambda$ centered on a level where $\varphi_k=\varphi^*$, with $q\ll\Lambda\ll N$. This trial vector has Rayleigh quotient $E_\mathrm{top}(\varphi^*)$ up to corrections of order $\Lambda^{-2}$ from the envelope and of order $\Lambda/N$ from the phase drift in it, both of which vanish as $N\to\infty$ with $\Lambda\propto N^{1/2}$. The same cut-off construction underlies the continuity in the flux of the spectrum of the almost-Mathieu operator, Harper's equation with a general coupling, taken as a union over phases~\cite{avron1990}. For smooth Hofstadter-like operators, Harper's model among them, the top and the bottom of the spectrum are Lipschitz continuous in the flux~\cite{bellissard1994}.

For the upper bound, the levels are covered by overlapping windows of width $\Lambda$ together with a smooth partition of unity, a set of window functions whose squares sum to one. For a tridiagonal matrix the expectation value in any vector equals the sum of the expectation values in the vector multiplied by each window function, up to a remainder of order $\Lambda^{-2}$ times the squared norm of the vector, with a coefficient depending only on the size of the off-diagonal entries. Within each window the matrix differs from a periodic one at a fixed phase by an amount of order $\Lambda/N$, and the largest eigenvalue of a finite piece of a periodic Jacobi matrix with non-negative off-diagonal entries does not exceed the band top. Therefore $\tau_1$ lies above $\max_\varphi E_\mathrm{top}(\varphi)$ by at most terms of order $\Lambda^{-2}$ and $\Lambda/N$. For the approximants of a root of unity the phase runs over a full period $\pi/q$ of $E_\mathrm{top}$ as the level runs over the $m$ values, and the maximum is thus attained in the interior.

The band top of the ring of Eq.~\eqref{eq:ring} is even in $\varphi$ and largest at $\varphi=\pi/(2q)$ for $q\geq3$ and finite $\varepsilon\neq2\sin\eta_0$. At $\varepsilon=2\sin\eta_0$ the parametrization by $s$ is singular, and the statement holds in the limit. To see this, expand $\det[E-H(\varphi)]$, with $H(\varphi)$ the matrix of the ring, over the covers of the ring by single sites and bonds. A term multiplying $E^k$ with $k\geq1$ covers $q-k$ sites and is a trigonometric polynomial of degree at most $2(q-k)$ in $\varphi$, while the period $\pi/q$ of the spectrum admits only the harmonics $e^{2iqn\varphi}$, so these coefficients do not depend on $\varphi$, by the argument used for the almost-Mathieu operator in Refs.~\cite{helffer1988harper,avron1990,jitomirskaya2022}. With the site phase $x_r=r\eta_0+\varphi$, for which $x_{r+1}$ is that of the next site, and the sign $\chi=\cos(2\,\mathrm{Re}\,\eta_0s)=\pm1$, the entries are $d_r=\cosh(2\,\mathrm{Im}\,\eta_0s)+\chi\cos2x_r$ and $b_r^2=\tfrac14[\cosh(4\,\mathrm{Im}\,\eta_0s)-\cos2x_r](1-\cos2x_{r+1})$. In the constant term the harmonics $e^{\pm2iq\varphi}$ come only from covers in which every site contributes the top harmonic $-(\chi/2)e^{2ix_r}$ of $-d_r$ and every bond the top harmonic $-\tfrac1{16}e^{2i(x_r+x_{r+1})}$ of $-b_r^2$. Summed over the covers of the cycle these give $\lambda_1^q+\lambda_2^q=2(-\chi/4)^q$, where $\lambda_{1,2}$ are the roots of $\lambda^2+(\chi/2)\lambda+\tfrac1{16}=(\lambda+\chi/4)^2$, times the phase $e^{i\eta_0q(q-1)}e^{2iq\varphi}=(-1)^{p(q-1)}e^{2iq\varphi}$. With the complex conjugate this is $K\cos(2q\varphi)$ with the real $K=4^{1-q}(-\chi)^q(-1)^{p(q-1)}$, and the two cyclic terms give $-2\prod_rb_r$. For a complex number $z$ the product formula $\prod_{r=0}^{q-1}\sin(z+r\pi p/q)=\pm2^{1-q}\sin(qz)$ holds, and it turns this product into $4^{1-q}|\sin q\varphi|(\sin^2q\varphi+S^2)^{1/2}$ with $S=\sinh(2q\,\mathrm{Im}\,\eta_0s)$. The top eigenvalue is a simple root and therefore increases with $\mathcal{G}(\varphi)=2|\sin q\varphi|(\sin^2q\varphi+S^2)^{1/2}-\sigma\cos(2q\varphi)$, where $\sigma=4^{q-1}K$. As a function of the variable $u=\sin^2q\varphi$, the derivative $(2u+S^2)/[u(u+S^2)]^{1/2}+2\sigma$ is positive, so the band top is even in $\varphi$, increases on $[0,\pi/(2q)]$ and peaks at $\varphi=\pi/(2q)$. Each step is checked numerically in the ten cases of Table~\ref{tab:nstar}. For $q=2$ the two bonds join the same pair of sites, so that the ring is the $2\times2$ matrix with the diagonal entries $d_0$ and $d_1$ of Eq.~\eqref{eq:ring} and, off the diagonal, the sum of its two bonds $b_0$ and $b_1$, whose top eigenvalue is $(d_0+d_1)/2+[(d_0-d_1)^2/4+(b_0+b_1)^2]^{1/2}$. The expansion above does not apply, and at the five couplings we checked this band top is even in $\varphi$ and largest at $\varphi=\pi/4$.

\section{Decay of eigenvectors and the crossover length}
\label{app:instanton}

The decay rate $\gamma$ of Sec.~\ref{sec:length} follows from the ring of Eq.~\eqref{eq:ring} with the site energies $d_r(\varphi)$ and the bonds $b_r(\varphi)$. A solution of the eigenvalue equation of the ring at the value $E$, continued periodically beyond one period, is multiplied by $e^{\pm q\gamma}$ over one period, where the discriminant $F(E,\varphi)$ of the ring fixes $\gamma$ through
\begin{equation}
2\cosh(q\gamma)=F(E,\varphi)\equiv\frac{\Pi_{0,q-1}(E)-b_{q-1}^2\,\Pi_{1,q-2}(E)}{\prod_{r=0}^{q-1}b_r} .
\label{eq:disc}
\end{equation}
Here the polynomial $\Pi_{0,q-1}(E)$, with leading coefficient one, is the characteristic polynomial of the open chain made of all $q$ sites of the ring, and $\Pi_{1,q-2}(E)$ is that of the chain without the two end sites, equal to one for $q=2$. The discriminant equals 2 at $E=E_\mathrm{top}(\varphi)$, and above this value $F>2$ and $\gamma$ is real and positive.

When the phase drifts slowly, $\varphi_k=k\delta$ with $|\delta|\ll1$, the detuning plays the role of Planck's constant, as the distance of the flux from a rational does for Harper's equation~\cite{helffer1990}, and an eigenvector of eigenvalue $E$ is, to exponential accuracy, $\exp(-\int\gamma\,dk)$ times a function with the period of the ring wherever $E>E_\mathrm{top}(\varphi_k)$. Between level 0 and the turning point, this gives Eq.~\eqref{eq:action_A} (the level sum turning into $1/|\delta|$ times a phase integral). The eigenvalues above $E_\mathrm{top}(0)$ are spaced by an amount of order $|\delta|$, and their contribution to $Z_n=\sum_j\psi_j(0)^2E_j^n$ is thus an integral of $\exp\{[n|\delta|\ln E-2W(E)]/|\delta|\}$ up to slowly varying factors. The saddle point of this integral is given by Eq.~\eqref{eq:saddle}, and the ratio $Z_n/Z_{n-1}$ is the value taken by $E$ at the saddle. Using $\partial\gamma/\partial E=\partial_EF/(q\sqrt{F^2-4})$, and since $\gamma$ vanishes at the turning point, the time $\Theta(E)$ of Eq.~\eqref{eq:saddle} becomes
\begin{equation}
\Theta(E)=2E\int_0^{\varphi_t(E)}\frac{\partial_EF(E,\varphi)}{q\sqrt{F(E,\varphi)^2-4}}\,d\varphi .
\label{eq:tau}
\end{equation}
Eigenvalues at or below $E_\mathrm{top}(0)$ come from eigenvectors that are not suppressed at level 0. They yield $E_\mathrm{top}(0)^n$ times a power of $n$, providing the leading contribution below the threshold, where the fraction $f$ is of order $1/m$ (Sec.~\ref{sec:length}).

Close to $\varphi=0$, the bond $b_{q-1}\simeq c_0|\varphi|$ in the denominator of Eq.~\eqref{eq:disc} makes $F$ large, and the turning point sits at $\varphi_t\simeq[E-E_\mathrm{top}(0)]/c$. For $E\to E_\mathrm{top}(0)$, the discriminant reduces to a function of the ratio $u=\varphi/\varphi_t$ alone, $F\simeq2/u$, hence $\partial_EF/\sqrt{F^2-4}\simeq1/(c\varphi_t\sqrt{1-u^2})$, which, inserted into Eq.~\eqref{eq:tau}, yields the threshold of Eq.~\eqref{eq:tauc}. We evaluated Eq.~\eqref{eq:tau} by quadrature with 30 digits. This result agrees with the difference quotient of $W(E)$ to a relative $\num{1.0e-13}$, and its limit agrees with Eq.~\eqref{eq:tauc} to $\num{7.5e-8}$ in the ten cases of Table~\ref{tab:nstar}. Across the 201 energies where it was computed for $1/3$ and $\varepsilon=1$, $\Theta(E)$ increases with $E$, as a maximum at the saddle requires. This behavior also holds for 19 crossing fractions in each of the ten cases, where each band top rises from $\varphi=0$ to the maximum. Here the prediction is sensitive to the ring. Altering the coupling from 2 to $2.2$ in the ring at $1/4$ shifts it \qty{14}{\percent} away from the calculated length. Four cases outside Table~\ref{tab:nstar} are compared to the formula in Table~\ref{tab:S5}.

\section{The momentum ring and the two actions}
\label{app:dual}

The matrix of the strong-driving ring is written as $VCV$, with the diagonal matrix $V$ of the $\sin(r\eta_0+\varphi)$ for $r=0,\dots,q-1$ and the circulant matrix $C$ having 2 on the diagonal and 1 between neighbors. Replacing the sines by their absolute values changes only signs removable by a diagonal transformation, since their product around the ring is positive. The matrix $VCV$ has the same nonzero eigenvalues as $C^{1/2}V^2C^{1/2}$. In the Fourier basis $C$ becomes diagonal with the weights $K_j=2+2\cos(2\pi j/q)=4\cos^2(\pi j/q)$, and
\begin{equation}
V^2=\tfrac12-\tfrac14\left(e^{2i\varphi}U+e^{-2i\varphi}U^\dagger\right),
\label{eq:S2}
\end{equation}
where the unitary matrix $U$, the multiplication by $e^{2ir\eta_0}$, shifts the momentum index $j$ by $p$. Since $p$ and $q$ are coprime, the shifts by $p$ connect all $q$ momenta in one ring, which has the site energies and bonds listed in Sec.~\ref{sec:strong}. Around this ring the product of the bond phases equals $(-1)^qe^{2iq\varphi}$. Ordering the sites on the ring so that site number $\nu$ has the momentum $\nu p$ modulo $q$ puts the weight $4\cos^2(\nu\eta_0)$ on that site. The momentum ring has the form of the original one with cosines in place of sines, and the model is self-dual.

Replacing $K_j$ with $K_j+4\mu^2$ in the deformed ring gives the symbol $\sin^2x\,[4\cos^2(\theta/2)+4\mu^2]$, which attains its maximum $4(1+\mu^2)$. Equating the symbol to this maximum and solving for the imaginary part of $x$ at real $\theta$, or of $\theta$ at real $x$, gives the two actions
\begin{equation}
\begin{aligned}
S_\theta(\mu)&=\int_0^{2\pi}\operatorname{arccosh}\sqrt{\frac{1+\mu^2}{\cos^2(\theta/2)+\mu^2}}\;d\theta,\\
S_x(\mu)&=\int_0^{\pi}\operatorname{arccosh}\left[\frac{2(1+\mu^2)}{\sin^2x}-1-2\mu^2\right]dx .
\end{aligned}
\label{eq:actions}
\end{equation}
At $\mu=0$ both reduce to $8G$, since for a real number $z$ not smaller than one $\operatorname{arccosh}(2z^2-1)=2\operatorname{arccosh}z$, and integration of $\operatorname{arccosh}(1/\cos\vartheta)$ over an angle $\vartheta$ from 0 to $\pi/2$ gives $2G$. With the commutator $[x,\theta]=i\eta$, the amplitude for one winding equals the exponential of minus the action divided by $\eta$, corresponding to a rate of the action divided by $\pi$ per unit of $q/p$.

\section{Validation against the Lindblad equation}
\label{si:validation}

\begin{figure}[htb]
\includegraphics[width=\columnwidth]{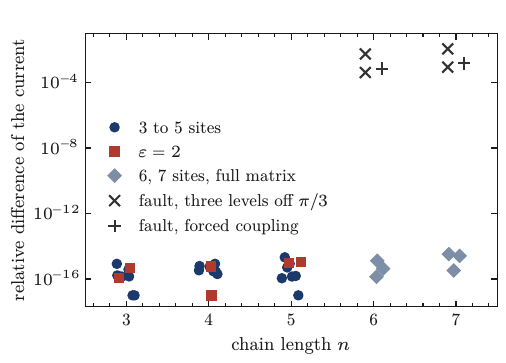}
\caption{\label{fig:S1}Validation of the matrix-product current against the steady state of the Lindblad equation solved directly, for chains of 3 to 7 sites. Blue circles denote the 21 cases of 3 to 5 sites. Red squares mark the six cases at $\varepsilon=2$. Blue-gray diamonds denote chains of 6 and 7 sites with the untruncated transfer matrix. The horizontal position within each chain length separates the cases. Crosses mark the three-level matrix away from $\eta=\pi/3$. Plus signs mark a coupling between the third and fourth levels forced at $\eta=\pi/3$, which the Lindblad steady state does not have. Differences that vanish in double precision are drawn at $10^{-17}$.}
\end{figure}

In the matrix-product steady state, the current agrees with the current obtained directly from the Lindblad equation without a matrix-product construction. In chains of 3 to 5 sites, the steady state was solved for explicitly as the generator null vector at 21 combinations of length, anisotropy angle, and coupling for $\varepsilon$ from $0.2$ to 5, and at six other combinations for $\varepsilon=2$, there also by a sparse direct solver. For the two currents, the largest relative difference is $\num{2.1e-15}$ (Fig.~\ref{fig:S1}).

Chains of 6 and 7 sites test the truncation at a root of unity. They are the shortest chains that can reach the fourth auxiliary level, so it matters here that the coupling between the third and fourth levels vanishes at $\eta=\pi/3$. At $\eta=\pi/3$ the three-level transfer matrix reproduces the current of the Lindblad steady state with a relative difference of at most $\num{2.6e-15}$. Away from this point, at $\eta=2\pi/7$ and $\eta=\pi/3+0.05$, the three-level matrix is off by up to \qty{1.2}{\percent}, and the untruncated matrix agrees to $\num{3.3e-15}$. The agreement resolves the truncation. A nonzero coupling between the third and fourth levels at $\eta=\pi/3$ would shift the current by a relative $\num{7.2e-4}$ for 6 sites and $\num{1.5e-3}$ for 7. Chains of 8 and 10 sites, which reach auxiliary levels 4 and 5, test the entries beyond the truncation of $\pi/3$. At $\eta=2\pi/7$, $\pi/3+0.05$ and $0.9$ with $\varepsilon=2$, the steady state solved directly gives the matrix-product current to a relative $\num{3.7e-15}$ at most. With the transfer matrix cut to the levels below the highest one the chain reaches, the same current misses by $\num{7.6e-5}$ to $\num{1.9e-4}$ for 8 sites and by $\num{7.9e-6}$ to $\num{3.2e-5}$ for 10 sites.

\section{Extrapolation of the approximants and the jumps at five roots of unity}
\label{si:jumps}

The limit of $J_\infty(l/m)$ for $l/m\to p/q$ is taken along sequences of approximants, and we use, where they exist, the sequences with $lq-pm=\pm1$ and $\pm2$ with denominators up to 4000. Each sequence is extrapolated with fits linear and quadratic in $1/m$ to the points with $m\geq200$ and with $m\geq400$, the limit of the sequence being taken as the mean of these fits. The quoted limit is the mean over the sequences, and the uncertainty is the larger of the range over the sequences and the largest spread of the fits within one sequence. Table~\ref{tab:S1} lists the relative jump $D$ obtained in this way at five roots of unity and three couplings, next to the value given by the band-top formula, Eq.~\eqref{eq:Dbt}. In all fifteen cases the two agree within the uncertainty of the extrapolation, the largest difference amounting to $\num{0.40}$ of the quoted uncertainty. The band-top value lies above the extrapolated one in 13 of the 15 cases, a bias of the extrapolation that stays within its quoted uncertainty. For three approximants the current was also computed with 100 digits, and double precision agrees with it to a relative $\num{1.0e-15}$.

The limit does not depend on the path, and there is no jump at an irrational target. For angles that are not roots of unity, with the transfer matrix kept to as many as $\num{1.2e6}$ levels, the lower branch is obtained from both sides of $\pi/3$, a quadratic extrapolation in the distance from $\pi/3$ yielding $J=\num{0.3607667}$ on either side at $\varepsilon=2$. The distances from $\pi/3$ are $\num{8.08e-4}$, $\num{4.08e-4}$, $\num{2.03e-4}$ and $\num{1.01e-4}$. At each the matrix is cut at between $\num{1.48e5}$ and $\num{1.19e6}$ levels, inversely proportional to the distance, and raising the cut by half changes the current by at most $\num{8.9e-14}$. A standard library bisection routine finds the top eigenvalue of the tridiagonal matrix, and linear and quadratic extrapolations give values of $D$ differing by at most $\num{2.2e-7}$. At $\varepsilon=2$ and with the matrix kept to $\num{400000}$ levels, a number that raised to $\num{600000}$ changes the current by $\num{1.4e-14}$, no jump appears at the irrational target $\sqrt2-1$, along whose continued-fraction convergents the current tends to the value at the target, differing from it by a relative $\num{2.9e-7}$ at $408/985$, $\num{6.0e-8}$ at $985/2378$ and $\num{9.0e-9}$ at $2378/5741$.

\begin{table}[htb]
\caption{\label{tab:S1}Relative jump $D$ in the current of the infinite chain for five roots of unity $\eta=\pi p/q$ at three bath couplings $\varepsilon$, from extrapolating approximants (uncertainty in the last digits) and from Eq.~\eqref{eq:Dbt}.}
\begin{ruledtabular}
\begin{tabular}{llS[table-format=2.5(2)]S[table-format=2.5]}
$p/q$ & $\varepsilon$ & {$D$ (\%), approximants} & {$D$ (\%), Eq.~\eqref{eq:Dbt}}\\
\hline
1/3 & 0.2 & 0.8070(4) & 0.80706\\
1/3 & 1 & 5.9090(11) & 5.90917\\
1/3 & 5 & 3.7459(23) & 3.74619\\
1/4 & 0.2 & 0.79450(12) & 0.79453\\
1/4 & 1 & 2.5109(4) & 2.51103\\
1/4 & 5 & 0.29995(12) & 0.29993\\
2/5 & 0.2 & 0.36686(25) & 0.36691\\
2/5 & 1 & 1.8519(6) & 1.85203\\
2/5 & 5 & 1.5407(12) & 1.54090\\
1/5 & 0.2 & 0.7286(5) & 0.72868\\
1/5 & 1 & 0.7916(14) & 0.79179\\
1/5 & 5 & 0.0446(10) & 0.04437\\
1/2 & 0.2 & 0.96149(12) & 0.96153\\
1/2 & 1 & 12.30465(26) & 12.30474\\
1/2 & 5 & 9.6963(6) & 9.69647\\
\end{tabular}
\end{ruledtabular}
\end{table}

\begin{figure}[htb]
\includegraphics[width=\columnwidth]{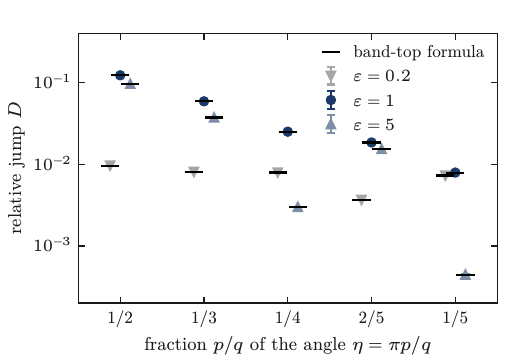}
\caption{\label{fig:S2}Relative jump $D$ of the infinite-chain current at five roots of unity (labeled $p/q$) for the bath couplings $\varepsilon=0.2$ (light gray down triangles, shifted left), 1 (blue circles) and 5 (blue-gray up triangles, shifted right), each from the extrapolation of up to four sequences of approximants with denominators up to 4000. Black bars show the band-top formula, Eq.~\eqref{eq:Dbt}. The uncertainties in Table~\ref{tab:S1} are smaller than the symbols.}
\end{figure}

\section{Dependence on the bath coupling}
\label{si:drive}

The approach to strong driving shown in Fig.~\ref{fig:drive} is documented in Table~\ref{tab:S2}. For the four odd denominators, the product $\varepsilon\,\Delta J$ of the coupling and the current jump converges to the jump of the strong-driving coefficient $\kappa$, the latter computed from the ring of the strong-driving limit with a different matrix. For the two even denominators the product decreases by a factor $10^{4}$ between $\varepsilon=100$ and $\varepsilon=10^4$, as $\varepsilon^{-2}$. The coefficient $\kappa$ itself is approached with corrections of relative order $\varepsilon^{-2}$. At $1/3$, the difference between $\kappa$ and $\varepsilon J_\infty$ is $\num{4.3e-4}$, $\num{4.3e-6}$ and $\num{4.3e-8}$ at $\varepsilon=100$, $10^3$ and $10^4$. The coupling $\varepsilon=2\sin\eta$, at which the complex parameter $s$ of Sec.~\ref{sec:model} is singular, is left out of this table. Among the listed couplings it occurs only for $p/q=1/2$ at $\varepsilon=2$, where Sec.~\ref{sec:drive} gives the limit $\varepsilon\to2$.

\begin{table*}[t]
\caption{\label{tab:S2}Approach to strong driving at six roots of unity. The product $\varepsilon\,\Delta J$ is the jump of the infinite-chain current multiplied by the coupling, from Eq.~\eqref{eq:Dbt}, and $\Delta\kappa$ is the jump of the strong-driving coefficient, Eq.~\eqref{eq:kjump}.}
\begin{ruledtabular}
\begin{tabular}{lS[table-format=1.3e-1]S[table-format=1.3e-1]S[table-format=1.3e-2]S[table-format=1.3e-1]}
$p/q$ & {$\varepsilon\,\Delta J$, $\varepsilon=1$} & {$\varepsilon\,\Delta J$, $\varepsilon=100$} & {$\varepsilon\,\Delta J$, $\varepsilon=10^4$} & {$\Delta\kappa$}\\
\hline
1/3 & 2.073e-2 & 3.954e-2 & 3.952e-2 & 3.952e-2\\
2/5 & 6.656e-3 & 1.749e-2 & 1.747e-2 & 1.747e-2\\
2/7 & 2.522e-3 & 8.961e-4 & 8.925e-4 & 8.925e-4\\
1/5 & 1.913e-3 & 1.022e-4 & 1.020e-4 & 1.020e-4\\
1/2 & 4.922e-2 & 7.984e-4 & 8.000e-8 & 0\\
1/4 & 7.395e-3 & 6.159e-6 & 6.161e-10 & 0\\
\end{tabular}
\end{ruledtabular}
\end{table*}

\section{Fits and prediction of the crossover length}
\label{si:crossover}

\begin{figure*}[t]
\includegraphics[width=\textwidth]{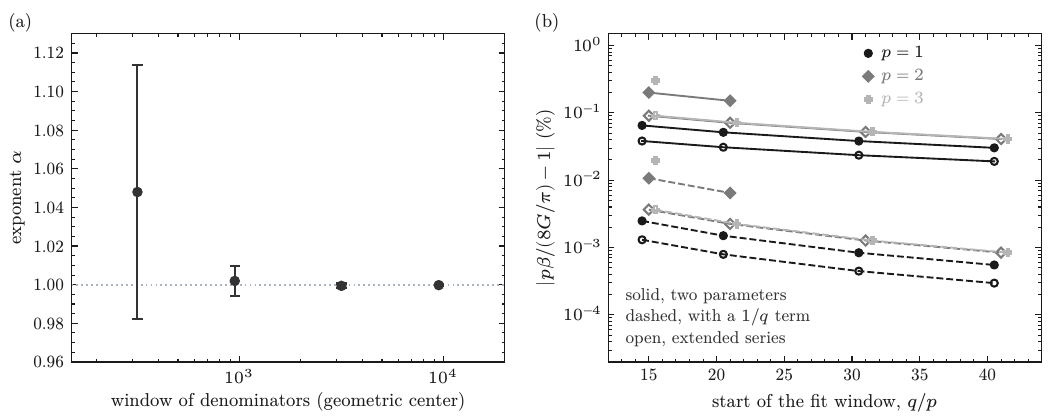}
\caption{\label{fig:S3}Fits. (a)~Exponent $\alpha$ of the crossover length, $n^*_f\propto m^\alpha$, in four windows of denominators. The mid-value and half-range over twelve fits are plotted at the geometric center of each window. The dotted line denotes $\alpha=1$. (b)~Absolute deviation from $8G/\pi$ of $p$ times the decay rate $\beta$ fitted to the jumps of $\kappa$ with the power of $q$ fixed at $-5/2$, against the smallest $q/p$ of the fit window, for the numerators $p=1$, 2 and 3 (key in the panel). Solid lines use two parameters, and dashed lines include a further term in $1/q$. Filled symbols give the values up to $q=81$ ($p=1$) and $q=61$ ($p=2$, 3), and open symbols include the recomputed extension to $q/p=60$ and to $q=121$ for $p=1$. The dotted line denotes zero deviation.}
\end{figure*}

The exponent of the crossover length converges to one as the window of denominators goes up. In each of the four windows $100\leq m\leq1000$, $300\leq m\leq3000$, $1000\leq m\leq10000$ and $3000\leq m\leq30000$ we take eight denominators on each side of $1/3$ that are spaced by a factor of about $\num{1.39}$ (3001, 4168, 5791, 8047, 11182, 15538, 21592 and 30001 below $1/3$ in the largest window), at $\varepsilon=1$ and 2, and fit $n^*_f\propto m^\alpha$ for $f=1/4$, $1/2$ and $3/4$, which amounts to twelve fits in all. The mid-value of these twelve exponents is $\num{1.048}$, $\num{1.0019}$, $\num{0.9994}$ and $\num{0.9998}$ in the four windows, and their half-range is $\num{0.066}$, $\num{0.0079}$, $\num{0.0012}$ and $\num{0.0004}$ [Fig.~\ref{fig:S3}(a)]. The crossover length is the first crossing of the fraction $f$ as interpolated linearly from one chain length to the next, because for small denominators this fraction is not monotonic in $n$.

To compute the crossover lengths, the original transfer matrix is iterated with every component stored as a mantissa and a binary exponent. In each multiplication by $T$, the three components entering a new component are brought to their largest exponent and added, with the sum renormalized to a mantissa between $1/2$ and 1. The current $J(n)=\zeta Z_{n-1}/Z_n$ is read from the first component of $T^ne_0$, where $e_0$ is the vector of level 0. Every entry of $T$ is nonnegative, so the iteration adds only nonnegative numbers and loses no digits by cancellation. Iterating the symmetrized matrix in logarithmic variables, with the current of the infinite chain found by bisection with Sturm sequences, gives for the exponent in the window $1000\leq m\leq10000$ the mid-value $\num{0.9994}$ with the half-range $\num{0.0012}$, as the iteration with stored exponents does. Iterating the symmetrized matrix with the squared norms of the iterates gives the same crossover lengths for all denominators up to 11183 and departs for $m\geq15538$, where those lengths exceed $10^5$ sites. There, components of the iterated vector fall below the range of double precision, and iterating the original matrix without an exponent for every component departs in the same way. The saddle-point formula in Sec.~\ref{sec:tunnel} matches the computed lengths at $m\approx30000$ to a relative $\num{1.7e-4}$.

The prediction of Sec.~\ref{sec:tunnel} holds away from the half-crossing too. Table~\ref{tab:S3} gives the crossover lengths for $f=1/4$ and $f=3/4$ at the five roots of unity and the two couplings of Table~\ref{tab:nstar}, along with the values of Eq.~\eqref{eq:nstar}. Apart from $1/5$ at $\varepsilon=2$, the mean of the two sides agrees with the prediction to \qty{0.2}{\percent} or better.

\begin{table*}[t]
\caption{\label{tab:S3}Crossover lengths $n^*_f/m$ for $f=1/4$ and $f=3/4$ at five roots of unity and two bath couplings, computed for the approximants with $|lq-pm|=1$ and $m\approx3000$ on the two sides of $p/q$, and predicted by Eq.~\eqref{eq:nstar}.}\begin{ruledtabular}
\begin{tabular}{llS[table-format=2.3]S[table-format=2.3]S[table-format=2.3]S[table-format=3.3]S[table-format=3.3]S[table-format=3.3]}
 & & \multicolumn{3}{c}{$f=1/4$} & \multicolumn{3}{c}{$f=3/4$}\\
\cline{3-5}\cline{6-8}
$p/q$ & $\varepsilon$ & {$l/m>p/q$} & {$l/m<p/q$} & {predicted} & {$l/m>p/q$} & {$l/m<p/q$} & {predicted}\\
\hline
1/3 & 1 & 4.823 & 4.816 & 4.817 & 6.956 & 6.949 & 6.950\\
1/3 & 2 & 5.247 & 5.236 & 5.238 & 7.309 & 7.297 & 7.299\\
1/4 & 1 & 9.284 & 9.249 & 9.258 & 12.519 & 12.477 & 12.487\\
1/4 & 2 & 18.099 & 17.935 & 17.988 & 22.932 & 22.667 & 22.769\\
2/5 & 1 & 9.670 & 9.652 & 9.654 & 13.898 & 13.880 & 13.879\\
2/5 & 2 & 9.071 & 9.040 & 9.048 & 13.027 & 12.996 & 13.001\\
2/7 & 1 & 16.258 & 16.114 & 16.168 & 23.269 & 23.127 & 23.172\\
2/7 & 2 & 22.061 & 21.549 & 21.777 & 31.259 & 30.758 & 30.968\\
1/5 & 1 & 24.985 & 24.686 & 24.787 & 31.503 & 31.028 & 31.212\\
1/5 & 2 & 94.588 & 87.483 & 90.953 & 116.346 & 105.911 & 110.104\\
\end{tabular}
\end{ruledtabular}
\end{table*}

The coupling $\varepsilon_\mathrm{tri}$ of Sec.~\ref{sec:tunnel} is located as the sign change, in the coupling, of $\Theta(E_f)-\Theta_c$, with $\Theta_c=\pi E_\mathrm{top}(0)/(qc)$ in closed form and $E_f$ the energy at a fraction $f$ of the jump, for $f=10^{-3}$, $3\times10^{-4}$ and $10^{-4}$, and extrapolated linearly to $f\to0$. The extrapolations from the two pairs of fractions differ by at most $\num{1.9e-6}$. Below $\varepsilon_\mathrm{tri}$, $\Theta$ increases on a grid of 19 fractions of the jump in every case of Table~\ref{tab:nstar}.

\section{Convergence and fits at strong driving, and the deformed ring}
\label{si:strong}

\begin{table*}[t]
\caption{\label{tab:S6}Actions of the deformed ring along the momentum and along the coordinate, divided by $\pi$, and the decay rate $\beta$ fitted to $\ln\Delta\kappa$ for odd $q$ in the windows given, with a term in $1/q$ and, in the last column, without it.}
\begin{ruledtabular}
\begin{tabular}{lS[table-format=1.4]S[table-format=1.4]S[table-format=1.4]S[table-format=1.4]S[table-format=1.4]S[table-format=1.4]}
$\mu$ & {$S_\theta/\pi$} & {$S_x/\pi$} & {$\beta$, 15--61} & {$\beta$, 21--61} & {$\beta$, 31--61} & {$\beta$, 31--61, no $1/q$}\\
\hline
0 & 2.3325 & 2.3325 & 2.3323 & 2.3324 & 2.3324 & 2.3313\\
0.25 & 1.8763 & 2.3713 & 1.8763 & 1.8762 & 1.8763 & 1.8752\\
0.5 & 1.5147 & 2.4772 & 1.5145 & 1.5146 & 1.5147 & 1.5138\\
1 & 1.0358 & 2.7986 & 1.0355 & 1.0356 & 1.0357 & 1.0354\\
\end{tabular}
\end{ruledtabular}
\end{table*}

\begin{table*}[t]
\caption{\label{tab:S5}The crossover formula of Sec.~\ref{sec:tunnel} for four cases outside Table~\ref{tab:nstar}. For each case, the two approximants on either side of the root of unity have the denominators $m$ listed. The predicted $n^*_f/m$ is that of Eq.~\eqref{eq:nstar}, and the computed value is the mean of the two sides. The last column is the relative deviation of the prediction from the computed mean, (predicted $-$ computed)/computed.}
\begin{ruledtabular}
\begin{tabular}{lllcS[table-format=2.3]S[table-format=2.3]S[table-format=-1.1e-1]}
$p/q$ & $\varepsilon$ & $m$ & $f$ & {predicted $n^*_f/m$} & {computed $n^*_f/m$} & {deviation}\\
\hline
1/3 & 0.5 & 3001, 3002 & 1/4 & 6.868 & 6.868 & 3.9e-5\\
1/3 & 0.5 & 3001, 3002 & 1/2 & 8.104 & 8.103 & 1.3e-4\\
1/3 & 0.5 & 3001, 3002 & 3/4 & 10.333 & 10.330 & 2.3e-4\\
1/3 & 5 & 3001, 3002 & 1/4 & 9.920 & 9.942 & -2.1e-3\\
1/3 & 5 & 3001, 3002 & 1/2 & 10.498 & 10.511 & -1.2e-3\\
1/3 & 5 & 3001, 3002 & 3/4 & 11.954 & 11.966 & -9.9e-4\\
3/7 & 1 & 3001, 3005 & 1/4 & 19.125 & 19.151 & -1.4e-3\\
3/7 & 1 & 3001, 3005 & 1/2 & 21.631 & 21.662 & -1.4e-3\\
3/7 & 1 & 3001, 3005 & 3/4 & 26.159 & 26.196 & -1.4e-3\\
3/8 & 1 & 3003, 3005 & 1/4 & 19.006 & 19.031 & -1.3e-3\\
3/8 & 1 & 3003, 3005 & 1/2 & 21.935 & 21.965 & -1.4e-3\\
3/8 & 1 & 3003, 3005 & 3/4 & 27.001 & 27.037 & -1.3e-3\\
\end{tabular}
\end{ruledtabular}
\end{table*}

\begin{figure}[htb]
\includegraphics[width=\columnwidth]{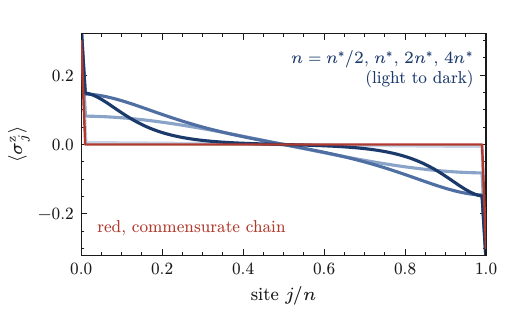}
\caption{\label{fig:S5}Magnetization $\langle\sigma^\mathrm{z}_j\rangle$ against $j/n$ at $1/3$ and $\varepsilon=1$ for the approximant $101/302$ with $n=n^*/2$, $n^*$, $2n^*$ and $4n^*$ (light to dark) and for the commensurate chain (red), which is unmagnetized away from its end sites.}
\end{figure}

\begin{figure*}[t]
\includegraphics[width=\textwidth]{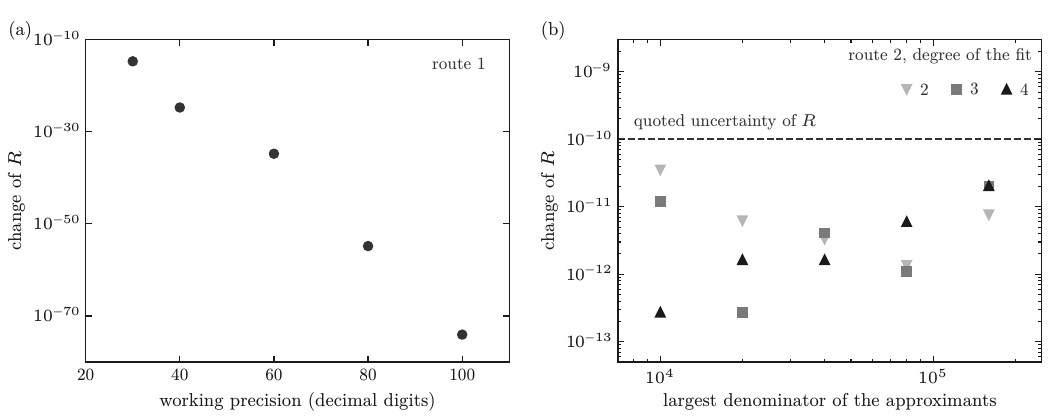}
\caption{\label{fig:S4}Stability of the ratio $R$ of the jumps of $\kappa$ at $1/9$ and $1/5$. (a)~First route, the change of $R$ from one working precision to the next, against the number of digits. (b)~Second route, the change of $R$ when the largest denominator of the approximants is doubled, against that denominator, for fits of degree 2 (light gray down triangles), 3 (gray squares) and 4 (black up triangles) in the inverse denominator (key in the panel). The dashed line denotes the uncertainty of $R$ quoted for this route. Both panels show the absolute change of $R$.}
\end{figure*}

The jumps of $\kappa$ are computed by two independent routes. The first route evaluates the ring of the strong-driving limit in high-precision arithmetic and scans the phase for the maximum of its top eigenvalue, with the ratio $R$ of the jumps at $1/9$ and $1/5$ changing by $\num{1.5e-55}$ from 60 to 80 digits and by $\num{8.6e-75}$ from 80 to 100 [Fig.~\ref{fig:S4}(a)]. The second route, working in double precision, extrapolates approximants from both sides with denominators up to $\num{1.6e5}$ and estimates the uncertainty of $R$ at $\num{1.0e-10}$ [Fig.~\ref{fig:S4}(b)]. Those two values for $R$ have a difference of $\num{2.4e-11}$, below this uncertainty. Finding each band top as the largest root of the characteristic polynomial of the ring reproduces the jumps of the first route where they overlap and extends the series (Table~\ref{tab:S4}). It evaluates the characteristic polynomial by continuants, with $100+1.5q/p$ digits, and finds its largest root by Newton's method started above all eigenvalues, from where the iteration converges to that root, or, for $p=1$, by bracketing and secant refinement. The working precision required grows with the denominator, rising from 66 digits at $q=5$ to 157 at $q=81$ in the first route, where 40 further digits change no jump by more than a relative $\num{9.2e-61}$, and to 281 digits at $q=121$ in the recomputation.

The fits of the decay law are stable against the window. Panel (b) of Fig.~\ref{fig:S3} gives $p$ times the fitted rate $\beta$ for the form $\ln\Delta\kappa=\ln A-\tfrac52\ln q-\beta q$ with a prefactor $A$, with and without a term in $1/q$, as the start of the window moves from $q/p=15$ to 41. With two parameters the deviation from $8G/\pi$ decreases from \qty{0.07}{\percent} to \qty{0.03}{\percent} for $p=1$ on the values up to $q=81$, and it is below \qty{0.02}{\percent} in each fit with the term in $1/q$. Allowing a free power of $q$ in the fits with the term in $1/q$, on the same windows, yields results between $\num{-2.50}$ and $\num{-2.55}$, lying closer to $-5/2$ the later the window starts. The wider range of Sec.~\ref{sec:tunnellaw} includes the fits without that term. Those that start at $q=11$ deviate from $8G/\pi$ by up to \qty{1.6}{\percent}.

\begin{table}[htb]
\caption{\label{tab:S4}Jumps $\Delta\kappa$ of the strong-driving coefficient at $\eta=\pi/q$ ($p=1$) and the local decay rate between consecutive odd denominators, $[\ln\Delta\kappa(q-2)-\ln\Delta\kappa(q)]/2$. The limit of the rate expected from the tunneling action is $8G/\pi=\num{2.33249}$. Values for $q\leq81$ are from the phase scan of the ring, those for $q>81$ from the characteristic polynomial. The table lists a selection of the series, and each local rate is computed from the two neighboring members of the full series.}
\begin{ruledtabular}
\begin{tabular}{lS[table-format=1.3e-3]S[table-format=1.4]}
$q$ & {$\Delta\kappa$} & {local rate}\\
\hline
3 & 3.952e-2 & \\
5 & 1.020e-4 & 2.9798\\
7 & 3.799e-7 & 2.7964\\
9 & 1.801e-9 & 2.6758\\
11 & 9.866e-12 & 2.6035\\
21 & 1.322e-22 & 2.4633\\
41 & 1.293e-43 & 2.3965\\
61 & 2.583e-64 & 2.3748\\
81 & 6.922e-85 & 2.3641\\
101 & 2.179e-105 & 2.3577\\
121 & 7.597e-126 & 2.3535\\
\end{tabular}
\end{ruledtabular}
\end{table}

In the deformed ring of Sec.~\ref{sec:tunnellaw}, the weights $K_j=4\cos^2(\pi j/q)$ of the momentum ring become $K_j+4\mu^2$, and in the coordinate basis it has diagonal $(2+4\mu^2)\sin^2(r\pi/q+\varphi)$ and bonds $|\sin(r\pi/q+\varphi)\sin[(r+1)\pi/q+\varphi]|$, for $p=1$. The jump is $4\sin^2(\pi/q)[1/\mathcal{E}(0)-1/\mathcal{E}(\pi/(2q))]$, computed for all odd $q$ from 5 to 61 with $50+1.3q$ digits and checked with 30 more, the value at $\varphi=\pi/(2q)$ being compared with a 12-point scan of the phase for $q\leq9$ and for $q\equiv1\pmod{10}$. Table~\ref{tab:S6} compares the actions with the rate $\beta$ from fits of $\ln\Delta\kappa=b+a\ln q-\beta q+e/q$, with free constants $b$, $a$ and $e$, over three windows and, in the last column, from that same form without the term in $1/q$.

\section{The lower bound on the Drude weight}
\label{si:bound}

To compute the bound $D_{\mathrm{Z}}$ given beside $\kappa$ in Sec.~\ref{sec:kappa}, one uses a tridiagonal matrix $B$ on the auxiliary levels 1 to $m-1$, with $D_{\mathrm{Z}}=1/[(1-B)^{-1}]_{11}$. In $B$, level $r$ has diagonal entry $\cos^2(r\eta)$, and between levels $r$ and $r+1$ the product of the two entries is $\sin^2(r\eta)\sin^2[(r+1)\eta]/4$. Thus the matrix $1-B$ has diagonal entries $\sin^2(r\eta)$ and shares the off-diagonal entry products of $M/2$, where $M$ denotes the matrix in Eq.~\eqref{eq:M}. This evaluation reproduces the closed-form values given in Ref.~\cite{prosen2011open}, among them $D_{\mathrm{Z}}=9/16$ at $\Delta=1/2$, at the five roots of unity where a comparison between the two was carried out.

The same matrix sets the current at weak coupling, Eq.~\eqref{eq:weak}. At $\varepsilon=0$ the entries of Eq.~\eqref{eq:T} on the levels $k\geq1$ are those of $2B$, namely the diagonal $2\cos^2(k\eta)$ and, across each bond, the product $\sin^2(k\eta)\sin^2[(k+1)\eta]$. Level 0 carries the diagonal entry $T_{0,0}=2|\cos(\eta s)|^2$ and couples to level 1 through the product $|\sin(2\eta s)|^2\sin^2\eta$ of the two entries that join them, which equals $\varepsilon^2$ up to terms of order $\varepsilon^4$. At a root of unity $1-B$ is positive definite, so the eigenvalues of $2B$ lie below 2, and second-order perturbation theory gives, up to terms of order $\varepsilon^4$, $\tau_1=T_{0,0}+\varepsilon^2[(2-2B)^{-1}]_{11}=T_{0,0}+\varepsilon^2/(2D_{\mathrm{Z}})$. With $\zeta=\varepsilon T_{0,0}/2$ and $T_{0,0}=2$ up to terms of order $\varepsilon^2$, Eq.~\eqref{eq:Jinf} yields Eq.~\eqref{eq:weak}. Along approximants the gap of $1-B$ closes as $m$ grows, so the limit of the jump is not taken from this expansion but from Eq.~\eqref{eq:Dbt}, evaluated for couplings down to $\num{1e-3}$ and extrapolated in $\varepsilon^2$. At $1/2$, $1/3$, $1/4$, $2/5$, $1/5$ and $2/7$ the result agrees with $\varepsilon^2/(2q\sin^2\eta_0)$, and approximants to $1/3$ at $\varepsilon=0.05$ give the same jump.

\section{Magnetization profiles}
\label{si:magnetization}

Inserting the Pauli matrix $\sigma^\mathrm{z}$ at site $j$ into the site trace of the transfer matrix in Eq.~\eqref{eq:T} yields a matrix $T^\mathrm{z}$ with zero diagonal and the bonds of $T$, taking a plus sign from level $k$ to $k+1$ and a minus sign from $k+1$ to $k$, so that $\langle\sigma^\mathrm{z}_j\rangle=\langle0|T^{j-1}T^\mathrm{z}T^{n-j}|0\rangle/Z_n$. This sign choice reproduces the null vector of the Lindblad generator for chains of 4 and 5 sites to $\num{2.6e-15}$, whereas reversing the signs misses it by $\num{1.3}$. For the long chains the vectors are iterated with 30 digits, and every profile is antisymmetric, $\langle\sigma^\mathrm{z}_j\rangle=-\langle\sigma^\mathrm{z}_{n+1-j}\rangle$, as the reflection of the chain combined with a spin flip requires. Figure~\ref{fig:S5} shows the profiles at $1/3$ and $\varepsilon=1$, and Sec.~\ref{sec:finitechain} quotes the sites $j=86$, 172 and 342, a tenth of the chains of 853, 1707 and 3413 sites.

\bibliography{refs}

\end{document}